\documentclass[aps,prl,reprint,superscriptaddress]{revtex4-2}
\usepackage{amsmath}
\usepackage{color}
\usepackage{soul}
\usepackage{bbm}
\usepackage{amssymb}
\usepackage{multirow}
\usepackage{amsbsy}
\usepackage{array}
\usepackage{diagbox}
\usepackage{bm}
\usepackage{extarrows}
\usepackage{graphicx}
\usepackage{lipsum}
\usepackage{bbding}
\usepackage{pifont}
\usepackage{makecell}
\usepackage[version=4]{mhchem}
\allowdisplaybreaks[4]
\usepackage[colorlinks=true,linkcolor=blue,citecolor=blue,urlcolor=blue,bookmarks=false]{hyperref}
\makeatletter
\let\savedaddcontentsline\addcontentsline
\newcommand{\disabletocentries}{%
  \let\addcontentsline\@gobblethree
}
\newcommand{\enabletocentries}{%
  \let\addcontentsline\savedaddcontentsline
}
\makeatother
\newcommand{\LNOtwoseven}{La$_3$Ni$_2$O$_7$}
\newcommand{\LNOfourten}{La$_4$Ni$_3$O$_{10}$}
\newcommand{\mub}{\mu_{\mathrm B}}
\newcommand{\bk}{\mathbf{k}}
\newcommand{\bR}{\mathbf{R}}
\newcommand{\bq}{\mathbf{q}}
\newcommand{\bQ}{\mathbf{Q}}
\newcommand{\ii}{\mathrm{i}}
\newcommand{\ee}{\mathrm{e}}
\newcommand{\RBZ}{\mathrm{RBZ}}

\begin{document}

\title{Itinerant Nature of Spin-Density-Wave Order in Ruddlesden–Popper Nickelates}

\author{Jiong Mei}
\affiliation{Beijing National Laboratory for Condensed Matter Physics and Institute of Physics, Chinese Academy of Sciences, Beijing 100190, China}
\affiliation{School of Physical Sciences, University of Chinese Academy of Sciences, Beijing 100190, China}

\author{Tianyang Xie}
\affiliation{Beijing National Laboratory for Condensed Matter Physics and Institute of Physics, Chinese Academy of Sciences, Beijing 100190, China}

\author{Kun Jiang}
\email{jiangkun@iphy.ac.cn}
\affiliation{Beijing National Laboratory for Condensed Matter Physics and Institute of Physics, Chinese Academy of Sciences, Beijing 100190, China}
\affiliation{School of Physical Sciences, University of Chinese Academy of Sciences, Beijing 100190, China}

\begin{abstract}
The nature of magnetism in layered Ruddlesden–Popper nickelates remains a central open question, particularly in light of recent observations of spin-wave-like magnetic excitations in metallic multilayer compounds. Here, we develop a unified itinerant description of spin-density-wave (SDW) order and magnetic excitations in \LNOtwoseven ~and \LNOfourten. The essential ingredient is the multilayer mirror structure of the NiO$_2$ blocks, which organizes the low-energy electronic states into mirror-even and mirror-odd sectors. We show that dominant interband nesting between mirror-opposite bands drives a mirror-selective itinerant SDW instability, whose collective modes naturally reproduce the experimentally observed spin-wave-like spectra. In \LNOfourten, the SDW further induces a secondary mirror-even charge density wave, yielding intertwined spin and charge textures. Our results demonstrate that magnetism in multilayer nickelates is fundamentally itinerant rather than local-moment in origin, and establish mirror-selective interband SDW order as a unifying organizing principle for magnetic correlations in these systems.
\end{abstract}

\maketitle

\disabletocentries

The discovery of superconductivity in layered Ruddlesden–Popper (RP) nickelates, beginning with La$_3$Ni$_2$O$_7$ under pressure \cite{Sun2023,2024Yuan,chengjg_crystal,zhangjunjie} and subsequently extending to related compounds La$_4$Ni$_3$O$_{10}$ \cite{4310-1,4310-2,4310-3,4310-4,4310-5}, has triggered intense interest in their intertwined spin, charge, and orbital degrees of freedom \cite{Review,2024Xiangb,2025Hwangc,2025Hwang,2025Chenb,2025Chenc}. Among the central open questions is the nature of magnetism in these materials and its potential connection to superconductivity \cite{wangmeng_sdw,taowu_327,shulei_PhysRevLett.132.256503,Chen2024,
Khasanov2025,taowu-4310,JPSJ.94.054704,RenXL2025,Luo_2025,chen2026naturemagnetismbilayernickelate,chen2026_4310,chan2026collectivespinexcitationstrilayer,
24f4-349n,PhysRevB.111.144502,PhysRevB.109.L180502,
Zhang2020,yajunyan_4310,caoyue_4310}.
Recent resonant inelastic x-ray scattering (RIXS) and neutron scattering experiments on both La$_3$Ni$_2$O$_7$ and La$_4$Ni$_3$O$_{10}$ have revealed dispersive magnetic excitations resembling conventional spin waves, and these spectra have often been analyzed using effective Heisenberg models \cite{Chen2024,chen2026naturemagnetismbilayernickelate,chen2026_4310,chan2026collectivespinexcitationstrilayer}. 
Such observations naturally suggest local-moment magnetism and strong superexchange interactions, drawing parallels to cuprates and other Mott-adjacent systems \cite{Chen2024,Yuxin_PhysRevB.110.205122,Botana_PhysRevMaterials.8.L111801,caokun_npj}.

However, this interpretation faces a fundamental difficulty. Unlike insulating cuprates, RP nickelates remain metallic over a wide temperature range and possess pronounced multiorbital itinerant electronic structures \cite{wangmeng_sdw,2024Yuan,4310-1,Zhang2020,
YangJG2024,lihaoxiang,yanglexian,yang2026electronicorigindensitywave,daweishen_4310}. 
Angle-resolved photoemission spectroscopy and first-principles calculations reveal multiple Fermi surface sheets with substantial hybridization across different orbitals and layers \cite{YangJG2024,lihaoxiang,yanglexian,yang2026electronicorigindensitywave,daweishen_4310,Yuxin_PhysRevB.110.205122}. 
Meanwhile, magnetic order, when observed, is often accompanied by relatively small ordered moments \cite{shulei_PhysRevLett.132.256503,taowu_327,taowu-4310,chen2026naturemagnetismbilayernickelate,Zhang2020}. 
Even more strikingly, fitting the measured excitation spectra within local-spin models often requires interlayer exchange couplings nearly an order of magnitude larger than the in-plane interactions, posing a serious challenge to a local-moment description \cite{Chen2024,chen2026naturemagnetismbilayernickelate,chen2026_4310,chan2026collectivespinexcitationstrilayer}. 
These observations raise a fundamental question: are the observed excitations genuine magnons of localized moments, or can apparently spin-wave-like spectra emerge from an underlying itinerant electronic system?

In this work, we show that magnetic excitations in RP nickelates can be naturally understood within a unified itinerant framework rooted in their multilayer electronic structures \cite{iron_PhysRevB.80.174401,iron_PhysRevB.81.140506,iron_PhysRevB.82.094441,iron_PhysRevB.83.224503,norman-landau}.
The key ingredient is the mirror symmetry of the layered NiO$_2$ blocks, which organizes low-energy electronic states into mirror-even and mirror-odd sectors \cite{Review}. We demonstrate that interband nesting between opposite mirror sectors drives a mirror-selective interband spin-density-wave instability, whose collective modes produce the experimentally observed magnon-like spectra. Applying this framework to both La$_3$Ni$_2$O$_7$ and La$_4$Ni$_3$O$_{10}$, we reproduce their distinct magnetic excitation spectra within a common mechanism. For La$_4$Ni$_3$O$_{10}$, we further show that the trilayer interband SDW naturally induces a secondary intraband charge density wave at $2\mathbf Q$, providing a microscopic route toward intertwined spin and charge order \cite{Zhang2020}. Our results demonstrate that spin-wave-like excitations in RP nickelates do not necessarily imply local-moment physics, and instead point toward an itinerant origin of magnetism in these multilayer nickelates.

\begin{figure}[!htbp]
\centering
\includegraphics[width=0.48\textwidth]{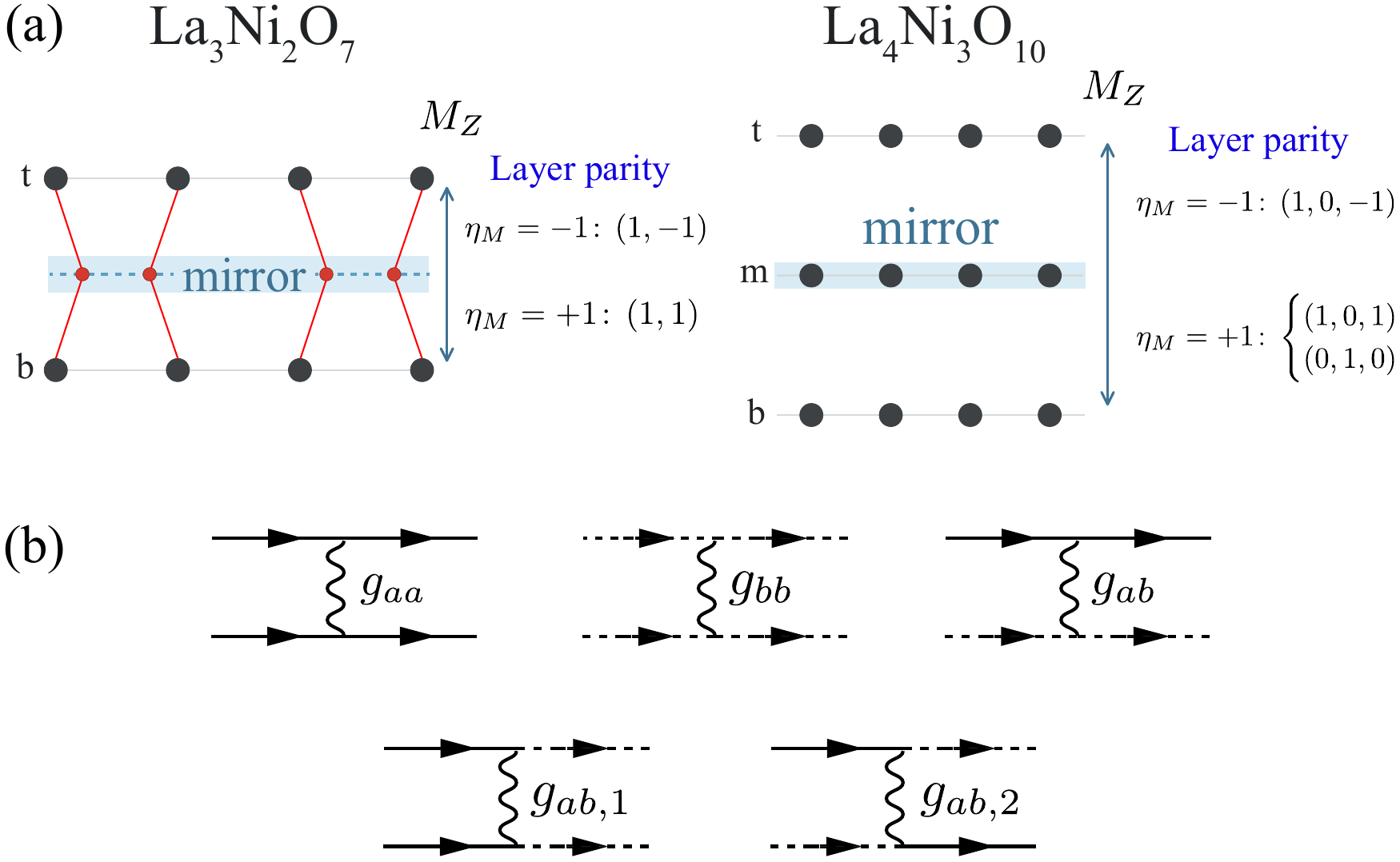}
\caption{Mirror-layer basis and effective interaction vertices.  (a) Schematic mirror-layer basis for \LNOtwoseven\ and \LNOfourten.  Black dots denote Ni sites in the NiO$_2$ layers, red dots indicate tilted apical oxygen positions along the vertical direction, and the shaded line marks the mirror plane.  The labels $t$, $m$, and $b$ denote top, middle, and bottom layers.  (b) Schematic Feynman diagrams for the five four-fermion vertices retained in the effective interaction $H_I^{ab}$.  Solid and dashed fermion lines denote the two retained bands $a$ and $b$, respectively \cite{iron_PhysRevB.80.174401}.  Momentum and spin labels are suppressed.}
\label{fig:mirror_layer_basis}
\end{figure}

Microscopically, the essential physics of RP nickelates originates from their quasi-two-dimensional NiO$_2$ layers. Given their itinerant electronic nature, it is natural to organize the low-energy degrees of freedom according to symmetry. Taking La$_3$Ni$_2$O$_7$ as an example, the low-pressure structure contains two NiO$_2$ layers and four Ni atoms per unit cell, as illustrated in Fig.~\ref{fig:mirror_layer_basis}(a). Under the mirror-$z$ operation, the top and bottom layers are exchanged, allowing the corresponding layer combinations to be classified into mirror-even states, $\eta_M=+1:(1,1)$, and mirror-odd states, $\eta_M=-1:(1,-1)$, where $(1,\pm1)$ denotes the relative phase between the two layers. In this basis, the noninteracting low-energy Hamiltonian can be written as
\begin{equation}
H_0=\sum_{\mathbf k,\lambda,s}\xi_\lambda(\mathbf k)c^\dagger_{\lambda\mathbf ks}c_{\lambda\mathbf ks},
\end{equation}
where $\lambda$ labels the low-energy bands, $s$ for electron spin and $\eta_\lambda=\pm1$ denotes their corresponding mirror eigenvalues.

\begin{figure}[!htbp]
\centering
\includegraphics[width=0.48\textwidth]{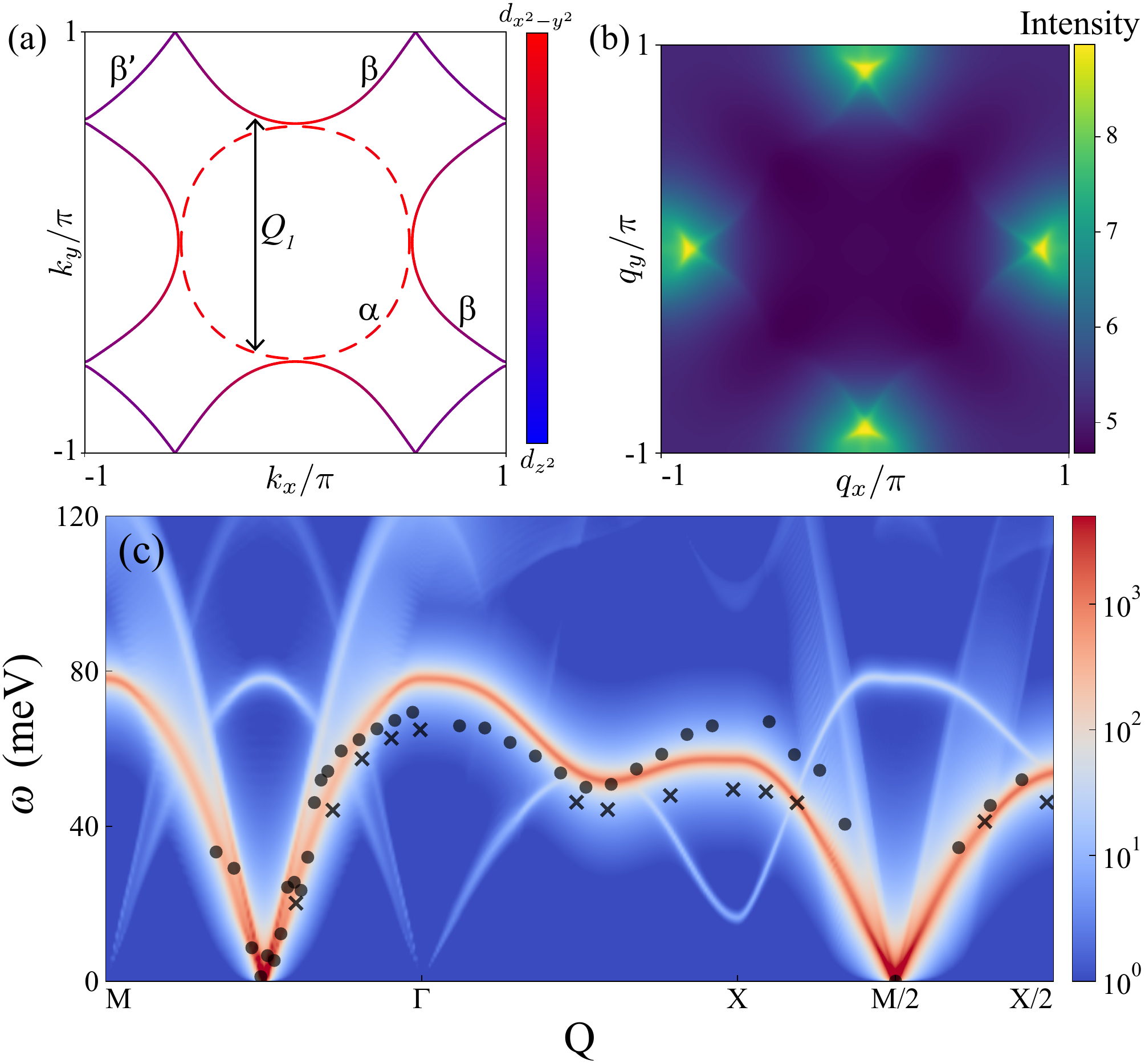}
\caption{Mirror-odd interband SDW and collective spin excitations in \LNOtwoseven.  (a) Mirror-resolved Fermi surface from the low-pressure tight-binding model taken from Ref.~\cite{Yuxin_PhysRevB.110.205122}.  Solid contours denote mirror-odd bands, while the dashed contour denotes the mirror-even band; the color scale indicates the orbital weight from $d_{z^2}$ to $d_{x^2-y^2}$.  The arrow $\mathbf Q_1$ marks the mirror-odd nesting channel retained in the main text.  (b) Normal-state interband bare susceptibility between $\alpha$ and $\beta$ bands, with dominant intensity near $\mathbf Q_1\simeq(0,\pm1.16\pi)$.  (c) Transverse spin spectral function calculated in the resulting commensurate interband SDW state, plotted along the high-pressure-phase Brillouin-zone path used for the high-symmetry labels.  The color bar in panel (c) gives the spectral intensity on a logarithmic scale.  Black dots denote the RIXS magnetic-excitation energies from Ref.~\cite{Chen2024}, and black crosses denote the inelastic neutron-scattering data from Ref.~\cite{chen2026naturemagnetismbilayernickelate}.}
\label{fig:lp327}
\end{figure}

In itinerant systems, magnetic order generally emerges from the leading spin instability of the Fermi surfaces (FSs). For \LNOtwoseven, we find that the dominant instability arises from interband scattering between mirror-even and mirror-odd bands, matching the symmetry of the experimentally observed SDW order ~\cite{Chen2024,chen2026naturemagnetismbilayernickelate,taowu_327}. 
The corresponding SDW order parameter can be schematically written as
\begin{equation}
\Delta_{ab}^{(\pm)}
\leftrightarrow
\sum_{\mathbf k}
\left\langle
c^\dagger_{a,\mathbf k\pm\mathbf Q,s}
\sigma^z_{ss'}
c_{b,\mathbf k,s'}
\right\rangle .
\end{equation}
where $\mathbf Q$ is the ordering wave vector, $a$ and $b$ label the nested bands connected by the dominant scattering channel.
For example, $a$ and $b$ correspond to the $\alpha$ and $\beta$ Fermi pockets shown in Fig.~\ref{fig:lp327}(a). 
For simplicity, we further assume that the magnetic moments are along the $z$ direction using the Pauli matrix $\sigma^z$.

To generate this ordered state, we supplement $H_0$ with minimal residual interactions within the retained low-energy sectors. Specifically, we retain only the interaction vertices necessary to capture the leading interband SDW instability and its associated collective spin excitations. The five relevant interaction channels are illustrated schematically in Fig.~\ref{fig:mirror_layer_basis}(b), where the solid and dashed fermion lines represent the two retained bands. The couplings $g_{aa}$ and $g_{bb}$ describe intraband density repulsion, $g_{ab}$ the interband density interaction, $g_{ab,1}$ pair transfer between the two bands, and $g_{ab,2}$ interband exchange~\cite{iron_PhysRevB.80.174401}. The corresponding residual interactions can be written as
\begin{equation}
\sum_{\langle ab\rangle}H_I^{ab}=\sum_{\langle ab\rangle} H_{aa}+H_{bb}+H_{ab}+H_{{\rm IT},1}+H_{{\rm IT},2}.
\end{equation}
Their explicit forms are given in the supplementary materials (SM) \cite{SuppMat}.

The transverse spin response is then computed about the resulting SDW Hartree-Fock (HF) ground state as
\begin{equation}
\chi^{+-}_{\rm RPA}(\mathbf q,\mathbf q';\omega)
=
\left[
\left(
\hat 1-\hat\chi^{+-}_{0,{\rm HF}}(\omega)\hat U^{+-}
\right)^{-1}
\hat\chi^{+-}_{0,{\rm HF}}(\omega)
\right]_{\mathbf q,\mathbf q'} .
\label{RPA}
\end{equation}
Here $\hat\chi^{+-}_{0,{\rm HF}}$ is the bare transverse susceptibility evaluated in the HF ordered state, and the hats denote matrices in band-pair and folded-momentum sectors.  The plotted spectral intensity is $I(\mathbf q,\omega)=\mathrm{Im}\,\chi^{+-}_{\rm phys}(\mathbf q,\omega)$, where $\chi^{+-}_{\rm phys}$ is obtained from $\chi^{+-}_{\rm RPA}(\mathbf q,\mathbf q';\omega)$ by projecting onto the physical transverse-spin operator and summing over the internal indices.  The detailed self-consistency equations, projection factors, and RPA vertex matrices $U^{+-}$ are left to the SM \cite{SuppMat}.

We also want to clarify the mirror convention adopted for \LNOfourten. Although the physical low-pressure structure only approximately preserves mirror-$z$ symmetry, a transparent low-energy description can still be constructed in a symmetry-adapted basis \cite{Review}. To this end, we use the three-layer high-pressure 4310 structure as an effective mirror-symmetric representation, while retaining parameters appropriate for the low-pressure phase. In this basis, the $M_z$ operation exchanges the two outer layers while leaving the middle layer unchanged, yielding one mirror-odd combination, $\eta_M=-1:(1,0,-1)$, and two mirror-even combinations, $\eta_M=+1:(1,0,1)$ and $(0,1,0)$, as shown in Fig.~\ref{fig:mirror_layer_basis}(a).

\textit{327-SDW}~
We first consider the SDW instability in \LNOtwoseven. As shown in Fig.~\ref{fig:lp327}(a), the Fermi surface consists of three sectors, labeled $\alpha$, $\beta$, and $\beta'$, with mirror eigenvalues $\eta_\alpha=+1$ and $\eta_{\beta/\beta'}=-1$. As discussed above, the dominant SDW instability originates from interband scattering between the $\alpha$ and $\beta$ pockets \cite{Chen2024,Yuxin_PhysRevB.110.205122}. Correspondingly, the normal-state interband susceptibility exhibits its strongest peak near $\mathbf{Q}_1 \simeq (0,\pm1.16\pi)$ or $(\pm1.16\pi,0)$, as shown in Fig.~\ref{fig:lp327}(b). In the HF/RPA calculation, we therefore adopt the nearby commensurate ordering vector $\mathbf{Q}=(0,\pi)$ to match experimental observations \cite{Chen2024}.

For the ordered-state calculation, we retain only the two nested bands, $\alpha$ and $\beta$, as the interacting subspace. The remaining bands are included in $H_0$ and the filling constraint but serve only as charge reservoirs, without residual interaction vertices. Within this minimal framework, only the interband density interaction $g_{\alpha\beta}$ and the pair-transfer interaction $g_{\alpha\beta,1}$ contribute to the SDW channel, while all other interaction terms are set to zero for simplicity. We further employ a globally renormalized tight-binding model to match the bandwidth observed in ARPES measurements~\cite{Yuxin_PhysRevB.110.205122,YangJG2024,SuppMat}.

Here, we take $g_{\alpha\beta}+g_{\alpha\beta,1}=0.405\,\mathrm{eV}$ and decouple the interactions into the SDW order parameters $\Delta_{\alpha\beta}$, followed by a self-consistent HF calculation. To visualize the ordered state in real space, we project the band-basis HF solution onto the Ni sublattices after fixing the Bloch-state phases according to their dominant mirror-resolved $d_{x^2-y^2}$ components, as detailed in the SM~\cite{SuppMat}. Within this convention, the resulting SDW state exhibits a strongly sublattice-selective magnetic pattern, with ordered moments approximately given by $\{0.39,0\}\,\mub$. 
The nearly vanishing moment on the second sublattice follows from destructive interference in the real-space projection: for the real $\mathbf Q=(0,\pi)$ SDW, the intracell phase at $\boldsymbol\tau_B=(1/2,1/2)$ makes the two folded-momentum components of the ordered density cancel in the local spin polarization. 
This site-selective magnetic pattern is consistent with the broader experimental phenomenology of spinless/spinful, or high-/low-moment stripe order reported in \LNOtwoseven ~\cite{Chen2024,taowu_327,chen2026naturemagnetismbilayernickelate,JPSJ.94.054704}. More importantly, the small ordered moments—on the order of $0.1\mu_B$ in both our theory and experiments—provide strong evidence for the itinerant nature of the SDW state, in sharp contrast to a conventional local-moment picture~\cite{Chen2024,chen2026naturemagnetismbilayernickelate,chen2026_4310,chan2026collectivespinexcitationstrilayer}.

We now turn to the spin dynamics of this SDW phase, obtained from the standard RPA formalism in Eq.~\eqref{RPA}.  As shown in Fig.~\ref{fig:lp327}(c), the calculated spectrum contains two collective branches with strongly unequal spectral weights: near $\Gamma$, the high-intensity branch shown in red is predominantly interband, whereas the weak branch visible in white is mainly intraband.  This intensity hierarchy reflects the interband character of the underlying SDW order, whose transverse collective motion couples most strongly to the interband spin channel.
The overall dispersion closely resembles the spin-wave-like form used to parameterize the RIXS measurements~\cite{Chen2024} and neutron scattering~\cite{chen2026naturemagnetismbilayernickelate}, including the pronounced dip of the high-energy branch along the $\Gamma$--$X$ direction.

In local-spin parametrizations of the magnetic spectrum, this high-energy $\Gamma$ mode is associated with a large interlayer exchange $J_\perp$ \cite{Chen2024}. In the present itinerant description, it instead appears as the high-energy collective branch of the mirror-odd interband SDW, rather than as a direct measure of a microscopic local-moment interlayer exchange. These results demonstrate that the same mirror-odd interband SDW responsible for the static spin--spinless texture also naturally captures the key features of the magnetic excitation spectrum, including the characteristic dispersion and intensity profile in the $\mathbf q$--$\omega$ plane.

\begin{figure}[!htbp]
\centering
\includegraphics[width=0.48\textwidth]{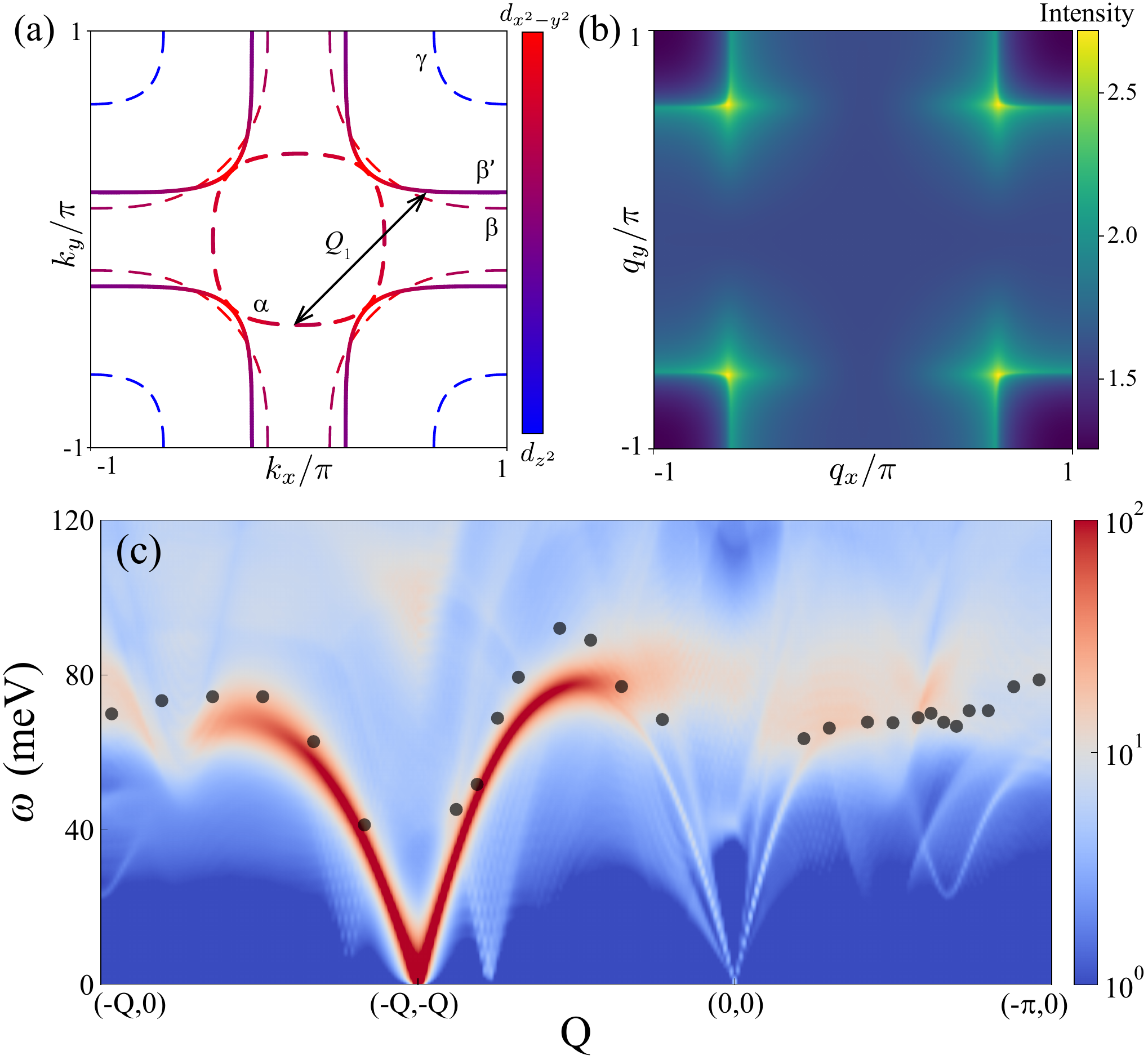}
\caption{Mirror-selective nesting and collective spin excitations in \LNOfourten.  (a) Mirror-resolved Fermi surface from the high-pressure-cell tight-binding model fitted to Ref.~\cite{yang2026electronicorigindensitywave}.  The color scale indicates the orbital weight from $d_{z^2}$ to $d_{x^2-y^2}$, and the arrow marks the nesting vector between the mirror-even $\alpha$ band and the mirror-odd $\beta^\prime$ band.  (b) Normal-state interband bare susceptibility in the $\alpha$--$\beta^\prime$ channel, with dominant intensity near $\mathbf Q_1\simeq(0.62\pi,0.62\pi)$.  (c) Transverse spin spectral function calculated in the commensurate interband SDW state with $\mathbf Q=(2\pi/3,2\pi/3)$. The color bar in panel (c) gives the spectral intensity on a logarithmic scale. Black dots denote RIXS magnetic-excitation energies from Ref.~\cite{chen2026_4310}.}
\label{fig:lp4310_spectrum}
\end{figure}

\textit{4310-SDW}~ 
We next apply the same strategy to the SDW order in \LNOfourten. Compared with \LNOtwoseven, the low-energy electronic structure contains four Fermi pockets, labeled $\alpha$, $\beta$, $\beta'$, and $\gamma$, as shown in Fig.~\ref{fig:lp4310_spectrum}(a) \cite{lihaoxiang,yang2026electronicorigindensitywave,daweishen_4310}. The dominant mirror-odd interband instability arises from scattering between the mirror-even $\alpha$ band and the mirror-odd $\beta^\prime$ band. Correspondingly, the interband susceptibility exhibits a pronounced peak near $\mathbf Q_1\simeq(0.62\pi,0.62\pi)$, as plotted in Fig.~\ref{fig:lp4310_spectrum}(b), consistent with the experimentally observed SDW ordering vector~\cite{Zhang2020,chen2026_4310,chan2026collectivespinexcitationstrilayer}. For the low-energy calculation, we also employ a globally renormalized tight-binding model adapted from Ref.~\cite{yang2026electronicorigindensitywave} for the low-pressure phase. Residual interactions are retained only within the nested $\alpha$–$\beta^\prime$ sector, while the remaining bands are treated as noninteracting spectator bands that contribute only to $H_0$ and the filling constraint. In the HF/RPA calculation, we approximate the ordering wave vector by the nearby commensurate value $\mathbf Q=(2\pi/3,2\pi/3)$.

For Fig.~\ref{fig:lp4310_spectrum}(c), we choose the interaction parameters as $g_{\alpha\alpha}=g_{\beta^\prime\beta^\prime}=0.24\,\mathrm{eV}$, $g_{\alpha\beta^\prime}=0.36\,\mathrm{eV}$, $g_{\alpha\beta^\prime,1}=0.24\,\mathrm{eV}$, and $g_{\alpha\beta^\prime,2}=0$. The resulting transverse RPA spectrum is noticeably broader and more diffuse than that of \LNOtwoseven. Nevertheless, the dominant branch near the ordering wave vector remains primarily interband in character and appears at an energy scale comparable to the magnetic excitations observed in RIXS experiments \cite{chan2026collectivespinexcitationstrilayer}. The overall spectral intensity is substantially weaker than in \LNOtwoseven, particularly along the $(0,0)$--$(-\pi,0)$ direction \cite{chan2026collectivespinexcitationstrilayer}.

Within the present itinerant framework, this reduced spectral weight naturally arises from stronger Landau damping in the more metallic trilayer system. Compared with \LNOtwoseven, the SDW gap in \LNOfourten\ is smaller, while several bands crossing the Fermi level remain outside the $\alpha$–$\beta^\prime$ ordered sector. As a result, quasiparticles participating in the SDW order can efficiently decay into ungapped low-energy states provided by these spectator bands, leading to a broadened collective response and a suppressed spectral intensity.

\begin{figure}[!htbp]
\centering
\includegraphics[width=0.48\textwidth]{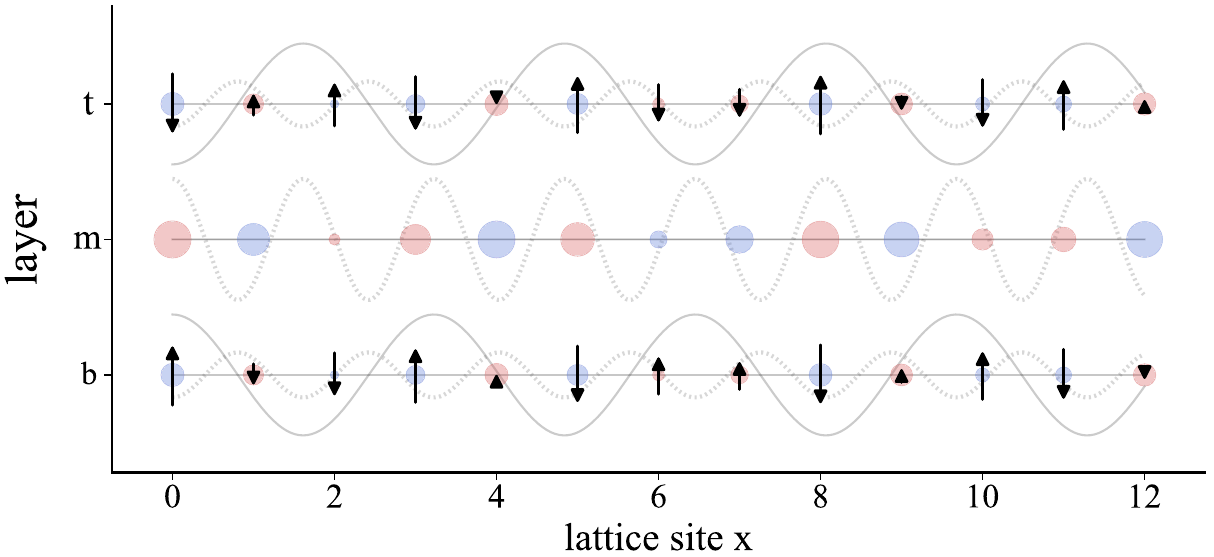}
\caption{Layer-resolved structure of the mirror-odd SDW and induced mirror-even $2\mathbf Q$ CDW in \LNOfourten\ for the same Hartree-Fock solution as in Fig.~\ref{fig:lp4310_spectrum}(c).  Black arrows denote the sign and relative magnitude of the spin modulation at $\mathbf Q$, while red and blue circles denote opposite signs of the charge modulation at $2\mathbf Q$, with circle size indicating the relative amplitude.  The solid and dotted gray curves are guides to the $\mathbf Q$ spin harmonic and the $2\mathbf Q$ charge harmonic, respectively.  The outer-layer moments reverse under mirror reflection and the middle-layer moment is strongly suppressed, whereas the charge modulation is mirror even.}
\label{fig:lp4310_pattern}
\end{figure}

Figure~\ref{fig:lp4310_pattern} summarizes the layer-resolved spin and charge textures obtained from the same HF solution. Projecting the ordered state onto the orbital basis using the mirror-resolved $d_{x^2-y^2}$ convention introduced above, we obtain a mirror-odd spin configuration: the two outer layers carry opposite spin polarizations with local moments of approximately $0.11\,\mu_B$, while the inner layer remains nonmagnetic. This pattern is consistent with the magnetic structures inferred from neutron and RIXS measurements~\cite{Zhang2020,chen2026_4310,chan2026collectivespinexcitationstrilayer}.

Unlike the case of \LNOtwoseven, the SDW order can induce a genuine second-harmonic charge modulation rather than merely a folded uniform component. In the band basis, the induced CDW corresponds to spin-even intraband coherences
\begin{equation}
\rho_\lambda(2\mathbf Q)
\leftrightarrow
\sum_{\mathbf k,s}
\left\langle
c^\dagger_{\lambda,\mathbf k+2\mathbf Q,s}
c_{\lambda,\mathbf k,s}
\right\rangle ,
\end{equation}
with $\lambda=\alpha,\beta^\prime$. Since these are intraband charge coherences, their mirror parity is given by $\eta_\lambda^2=+1$, implying that the induced $2\mathbf Q$ CDW is mirror even.

Projecting back to the layer basis, we find that the charge modulation is strongest on the inner layer, with an amplitude of approximately $0.023$, while the two outer layers exhibit weaker modulations of equal magnitude, about $8.6\times10^{-3}$, as illustrated in Fig.~\ref {fig:lp4310_pattern}. This provides a microscopic realization, in the mirror-band basis, of the intertwined SDW–CDW order allowed by Landau theory~\cite{norman-landau}.

In summary, we have developed a unified itinerant description of spin-density-wave order and magnetic excitations in layered RP nickelates. The central ingredient is the multilayer mirror structure of the NiO$_2$ blocks, which naturally organizes the low-energy electronic states into mirror-even and mirror-odd sectors. Within this framework, the dominant magnetic instability originates from mirror-odd interband scattering between nested Fermi-surface pockets. Applying this mechanism to both \LNOtwoseven~ and \LNOfourten, we show that the same itinerant SDW framework can simultaneously account for the experimentally observed ordering wave vectors, spin textures, and spin-wave-like magnetic excitation spectra. For \LNOfourten, the SDW also generates a secondary mirror-even charge modulation, providing a route toward intertwined SDW–CDW order.

More importantly, our results demonstrate that magnetism in multilayer nickelates is fundamentally itinerant rather than local-moment in origin. The small ordered moments arise directly from Fermi-surface reconstruction within a partially metallic multiband system, while the magnetic excitations correspond to collective interband modes instead of conventional magnons. Consequently, the large effective exchange couplings extracted from phenomenological spin-wave fits should not be viewed as microscopic superexchange interactions. Indeed, within a local-moment picture, a sufficiently large interlayer $J_\perp$ would favor interlayer valence-bond singlets rather than magnetic order as demonstrated in SM \cite{SuppMat}, further supporting the itinerant origin of the SDW state.

Furthermore, although we employ globally renormalized tight-binding models to approximately reproduce the Fermi-surface topology and energy scales, the essential itinerant SDW physics does not depend on these details. The key ingredients are the multilayer mirror structure and the dominant interband nesting between mirror-opposite sectors, which establish mirror-selective interband SDW order as a unifying principle for magnetism in multilayer nickelates.

\textit{Acknowledgement}
We acknowledge the support by the National Natural Science Foundation of China (Grant NSFC-12494594), the Chinese Academy of Sciences Project for Young Scientists in Basic Research (2022YSBR-048), and the New Cornerstone Investigator Program. 

\bibliographystyle{apsrev4-2}
\bibliography{references}

\begin{thebibliography}{48}%
\makeatletter
\providecommand \@ifxundefined [1]{%
 \@ifx{#1\undefined}
}%
\providecommand \@ifnum [1]{%
 \ifnum #1\expandafter \@firstoftwo
 \else \expandafter \@secondoftwo
 \fi
}%
\providecommand \@ifx [1]{%
 \ifx #1\expandafter \@firstoftwo
 \else \expandafter \@secondoftwo
 \fi
}%
\providecommand \natexlab [1]{#1}%
\providecommand \enquote  [1]{``#1''}%
\providecommand \bibnamefont  [1]{#1}%
\providecommand \bibfnamefont [1]{#1}%
\providecommand \citenamefont [1]{#1}%
\providecommand \href@noop [0]{\@secondoftwo}%
\providecommand \href [0]{\begingroup \@sanitize@url \@href}%
\providecommand \@href[1]{\@@startlink{#1}\@@href}%
\providecommand \@@href[1]{\endgroup#1\@@endlink}%
\providecommand \@sanitize@url [0]{\catcode `\\12\catcode `\$12\catcode `\&12\catcode `\#12\catcode `\^12\catcode `\_12\catcode `\%12\relax}%
\providecommand \@@startlink[1]{}%
\providecommand \@@endlink[0]{}%
\providecommand \url  [0]{\begingroup\@sanitize@url \@url }%
\providecommand \@url [1]{\endgroup\@href {#1}{\urlprefix }}%
\providecommand \urlprefix  [0]{URL }%
\providecommand \Eprint [0]{\href }%
\providecommand \doibase [0]{https://doi.org/}%
\providecommand \selectlanguage [0]{\@gobble}%
\providecommand \bibinfo  [0]{\@secondoftwo}%
\providecommand \bibfield  [0]{\@secondoftwo}%
\providecommand \translation [1]{[#1]}%
\providecommand \BibitemOpen [0]{}%
\providecommand \bibitemStop [0]{}%
\providecommand \bibitemNoStop [0]{.\EOS\space}%
\providecommand \EOS [0]{\spacefactor3000\relax}%
\providecommand \BibitemShut  [1]{\csname bibitem#1\endcsname}%
\let\auto@bib@innerbib\@empty
\bibitem [{\citenamefont {Sun}\ \emph {et~al.}(2023)\citenamefont {Sun}, \citenamefont {Huo}, \citenamefont {Hu}, \citenamefont {Li}, \citenamefont {Liu}, \citenamefont {Han}, \citenamefont {Tang}, \citenamefont {Mao}, \citenamefont {Yang}, \citenamefont {Wang}, \citenamefont {Cheng}, \citenamefont {Yao}, \citenamefont {Zhang},\ and\ \citenamefont {Wang}}]{Sun2023}%
  \BibitemOpen
  \bibfield  {author} {\bibinfo {author} {\bibfnamefont {H.}~\bibnamefont {Sun}}, \bibinfo {author} {\bibfnamefont {M.}~\bibnamefont {Huo}}, \bibinfo {author} {\bibfnamefont {X.}~\bibnamefont {Hu}}, \bibinfo {author} {\bibfnamefont {J.}~\bibnamefont {Li}}, \bibinfo {author} {\bibfnamefont {Z.}~\bibnamefont {Liu}}, \bibinfo {author} {\bibfnamefont {Y.}~\bibnamefont {Han}}, \bibinfo {author} {\bibfnamefont {L.}~\bibnamefont {Tang}}, \bibinfo {author} {\bibfnamefont {Z.}~\bibnamefont {Mao}}, \bibinfo {author} {\bibfnamefont {P.}~\bibnamefont {Yang}}, \bibinfo {author} {\bibfnamefont {B.}~\bibnamefont {Wang}}, \bibinfo {author} {\bibfnamefont {J.}~\bibnamefont {Cheng}}, \bibinfo {author} {\bibfnamefont {D.-X.}\ \bibnamefont {Yao}}, \bibinfo {author} {\bibfnamefont {G.-M.}\ \bibnamefont {Zhang}},\ and\ \bibinfo {author} {\bibfnamefont {M.}~\bibnamefont {Wang}},\ }\href {https://doi.org/10.1038/s41586-023-06408-7} {\bibfield  {journal} {\bibinfo  {journal} {Nature}\ }\textbf {\bibinfo {volume} {621}},\ \bibinfo
  {pages} {493} (\bibinfo {year} {2023})}\BibitemShut {NoStop}%
\bibitem [{\citenamefont {Zhang}\ \emph {et~al.}(2024)\citenamefont {Zhang}, \citenamefont {Su}, \citenamefont {Huang}, \citenamefont {Shan}, \citenamefont {Sun}, \citenamefont {Huo}, \citenamefont {Ye}, \citenamefont {Zhang}, \citenamefont {Yang}, \citenamefont {Xu}, \citenamefont {Su}, \citenamefont {Li}, \citenamefont {Smidman}, \citenamefont {Wang}, \citenamefont {Jiao},\ and\ \citenamefont {Yuan}}]{2024Yuan}%
  \BibitemOpen
  \bibfield  {author} {\bibinfo {author} {\bibfnamefont {Y.}~\bibnamefont {Zhang}}, \bibinfo {author} {\bibfnamefont {D.}~\bibnamefont {Su}}, \bibinfo {author} {\bibfnamefont {Y.}~\bibnamefont {Huang}}, \bibinfo {author} {\bibfnamefont {Z.}~\bibnamefont {Shan}}, \bibinfo {author} {\bibfnamefont {H.}~\bibnamefont {Sun}}, \bibinfo {author} {\bibfnamefont {M.}~\bibnamefont {Huo}}, \bibinfo {author} {\bibfnamefont {K.}~\bibnamefont {Ye}}, \bibinfo {author} {\bibfnamefont {J.}~\bibnamefont {Zhang}}, \bibinfo {author} {\bibfnamefont {Z.}~\bibnamefont {Yang}}, \bibinfo {author} {\bibfnamefont {Y.}~\bibnamefont {Xu}}, \bibinfo {author} {\bibfnamefont {Y.}~\bibnamefont {Su}}, \bibinfo {author} {\bibfnamefont {R.}~\bibnamefont {Li}}, \bibinfo {author} {\bibfnamefont {M.}~\bibnamefont {Smidman}}, \bibinfo {author} {\bibfnamefont {M.}~\bibnamefont {Wang}}, \bibinfo {author} {\bibfnamefont {L.}~\bibnamefont {Jiao}},\ and\ \bibinfo {author} {\bibfnamefont {H.}~\bibnamefont {Yuan}},\ }\href
  {https://doi.org/10.1038/s41567-024-02515-y} {\bibfield  {journal} {\bibinfo  {journal} {Nature Physics}\ }\textbf {\bibinfo {volume} {20}},\ \bibinfo {pages} {1269} (\bibinfo {year} {2024})}\BibitemShut {NoStop}%
\bibitem [{\citenamefont {Wang}\ \emph {et~al.}(2024{\natexlab{a}})\citenamefont {Wang}, \citenamefont {Wang}, \citenamefont {Shen}, \citenamefont {Hou}, \citenamefont {Luo}, \citenamefont {Ma}, \citenamefont {Yang}, \citenamefont {Shi}, \citenamefont {Dou}, \citenamefont {Feng}, \citenamefont {Yang}, \citenamefont {Shi}, \citenamefont {Ren}, \citenamefont {Ma}, \citenamefont {Yang}, \citenamefont {Liu}, \citenamefont {Liu}, \citenamefont {Zhang}, \citenamefont {Dong}, \citenamefont {Wang}, \citenamefont {Jiang}, \citenamefont {Hu}, \citenamefont {Nagasaki}, \citenamefont {Kitagawa}, \citenamefont {Calder}, \citenamefont {Yan}, \citenamefont {Sun}, \citenamefont {Wang}, \citenamefont {Zhou}, \citenamefont {Uwatoko},\ and\ \citenamefont {Cheng}}]{chengjg_crystal}%
  \BibitemOpen
  \bibfield  {author} {\bibinfo {author} {\bibfnamefont {N.}~\bibnamefont {Wang}}, \bibinfo {author} {\bibfnamefont {G.}~\bibnamefont {Wang}}, \bibinfo {author} {\bibfnamefont {X.}~\bibnamefont {Shen}}, \bibinfo {author} {\bibfnamefont {J.}~\bibnamefont {Hou}}, \bibinfo {author} {\bibfnamefont {J.}~\bibnamefont {Luo}}, \bibinfo {author} {\bibfnamefont {X.}~\bibnamefont {Ma}}, \bibinfo {author} {\bibfnamefont {H.}~\bibnamefont {Yang}}, \bibinfo {author} {\bibfnamefont {L.}~\bibnamefont {Shi}}, \bibinfo {author} {\bibfnamefont {J.}~\bibnamefont {Dou}}, \bibinfo {author} {\bibfnamefont {J.}~\bibnamefont {Feng}}, \bibinfo {author} {\bibfnamefont {J.}~\bibnamefont {Yang}}, \bibinfo {author} {\bibfnamefont {Y.}~\bibnamefont {Shi}}, \bibinfo {author} {\bibfnamefont {Z.}~\bibnamefont {Ren}}, \bibinfo {author} {\bibfnamefont {H.}~\bibnamefont {Ma}}, \bibinfo {author} {\bibfnamefont {P.}~\bibnamefont {Yang}}, \bibinfo {author} {\bibfnamefont {Z.}~\bibnamefont {Liu}}, \bibinfo {author} {\bibfnamefont {Y.}~\bibnamefont
  {Liu}}, \bibinfo {author} {\bibfnamefont {H.}~\bibnamefont {Zhang}}, \bibinfo {author} {\bibfnamefont {X.}~\bibnamefont {Dong}}, \bibinfo {author} {\bibfnamefont {Y.}~\bibnamefont {Wang}}, \bibinfo {author} {\bibfnamefont {K.}~\bibnamefont {Jiang}}, \bibinfo {author} {\bibfnamefont {J.}~\bibnamefont {Hu}}, \bibinfo {author} {\bibfnamefont {S.}~\bibnamefont {Nagasaki}}, \bibinfo {author} {\bibfnamefont {K.}~\bibnamefont {Kitagawa}}, \bibinfo {author} {\bibfnamefont {S.}~\bibnamefont {Calder}}, \bibinfo {author} {\bibfnamefont {J.}~\bibnamefont {Yan}}, \bibinfo {author} {\bibfnamefont {J.}~\bibnamefont {Sun}}, \bibinfo {author} {\bibfnamefont {B.}~\bibnamefont {Wang}}, \bibinfo {author} {\bibfnamefont {R.}~\bibnamefont {Zhou}}, \bibinfo {author} {\bibfnamefont {Y.}~\bibnamefont {Uwatoko}},\ and\ \bibinfo {author} {\bibfnamefont {J.}~\bibnamefont {Cheng}},\ }\href {https://doi.org/10.1038/s41586-024-07996-8} {\bibfield  {journal} {\bibinfo  {journal} {Nature}\ }\textbf {\bibinfo {volume} {634}},\ \bibinfo
  {pages} {579} (\bibinfo {year} {2024}{\natexlab{a}})}\BibitemShut {NoStop}%
\bibitem [{\citenamefont {Li}\ \emph {et~al.}(2026)\citenamefont {Li}, \citenamefont {Xing}, \citenamefont {Peng}, \citenamefont {Dou}, \citenamefont {Guo}, \citenamefont {Ma}, \citenamefont {Zhang}, \citenamefont {Wang}, \citenamefont {Luo}, \citenamefont {Yang}, \citenamefont {Zhang}, \citenamefont {Chang}, \citenamefont {Chen}, \citenamefont {Cai}, \citenamefont {Cheng}, \citenamefont {Wang}, \citenamefont {Liu}, \citenamefont {Luo}, \citenamefont {Hirao}, \citenamefont {Matsuoka}, \citenamefont {Kadobayashi}, \citenamefont {Zeng}, \citenamefont {Zheng}, \citenamefont {Zhou}, \citenamefont {Zeng}, \citenamefont {Tao},\ and\ \citenamefont {Zhang}}]{zhangjunjie}%
  \BibitemOpen
  \bibfield  {author} {\bibinfo {author} {\bibfnamefont {F.}~\bibnamefont {Li}}, \bibinfo {author} {\bibfnamefont {Z.}~\bibnamefont {Xing}}, \bibinfo {author} {\bibfnamefont {D.}~\bibnamefont {Peng}}, \bibinfo {author} {\bibfnamefont {J.}~\bibnamefont {Dou}}, \bibinfo {author} {\bibfnamefont {N.}~\bibnamefont {Guo}}, \bibinfo {author} {\bibfnamefont {L.}~\bibnamefont {Ma}}, \bibinfo {author} {\bibfnamefont {Y.}~\bibnamefont {Zhang}}, \bibinfo {author} {\bibfnamefont {L.}~\bibnamefont {Wang}}, \bibinfo {author} {\bibfnamefont {J.}~\bibnamefont {Luo}}, \bibinfo {author} {\bibfnamefont {J.}~\bibnamefont {Yang}}, \bibinfo {author} {\bibfnamefont {J.}~\bibnamefont {Zhang}}, \bibinfo {author} {\bibfnamefont {T.}~\bibnamefont {Chang}}, \bibinfo {author} {\bibfnamefont {Y.-S.}\ \bibnamefont {Chen}}, \bibinfo {author} {\bibfnamefont {W.}~\bibnamefont {Cai}}, \bibinfo {author} {\bibfnamefont {J.}~\bibnamefont {Cheng}}, \bibinfo {author} {\bibfnamefont {Y.}~\bibnamefont {Wang}}, \bibinfo {author} {\bibfnamefont
  {Y.}~\bibnamefont {Liu}}, \bibinfo {author} {\bibfnamefont {T.}~\bibnamefont {Luo}}, \bibinfo {author} {\bibfnamefont {N.}~\bibnamefont {Hirao}}, \bibinfo {author} {\bibfnamefont {T.}~\bibnamefont {Matsuoka}}, \bibinfo {author} {\bibfnamefont {H.}~\bibnamefont {Kadobayashi}}, \bibinfo {author} {\bibfnamefont {Z.}~\bibnamefont {Zeng}}, \bibinfo {author} {\bibfnamefont {Q.}~\bibnamefont {Zheng}}, \bibinfo {author} {\bibfnamefont {R.}~\bibnamefont {Zhou}}, \bibinfo {author} {\bibfnamefont {Q.}~\bibnamefont {Zeng}}, \bibinfo {author} {\bibfnamefont {X.}~\bibnamefont {Tao}},\ and\ \bibinfo {author} {\bibfnamefont {J.}~\bibnamefont {Zhang}},\ }\href {https://doi.org/10.1038/s41586-025-09954-4} {\bibfield  {journal} {\bibinfo  {journal} {Nature}\ }\textbf {\bibinfo {volume} {649}},\ \bibinfo {pages} {871} (\bibinfo {year} {2026})}\BibitemShut {NoStop}%
\bibitem [{\citenamefont {Zhu}\ \emph {et~al.}(2024)\citenamefont {Zhu}, \citenamefont {Peng}, \citenamefont {Zhang}, \citenamefont {Pan}, \citenamefont {Chen}, \citenamefont {Chen}, \citenamefont {Ren}, \citenamefont {Liu}, \citenamefont {Hao}, \citenamefont {Li}, \citenamefont {Xing}, \citenamefont {Lan}, \citenamefont {Han}, \citenamefont {Wang}, \citenamefont {Jia}, \citenamefont {Wo}, \citenamefont {Gu}, \citenamefont {Gu}, \citenamefont {Ji}, \citenamefont {Wang}, \citenamefont {Gou}, \citenamefont {Shen}, \citenamefont {Ying}, \citenamefont {Chen}, \citenamefont {Yang}, \citenamefont {Cao}, \citenamefont {Zheng}, \citenamefont {Zeng}, \citenamefont {Guo},\ and\ \citenamefont {Zhao}}]{4310-1}%
  \BibitemOpen
  \bibfield  {author} {\bibinfo {author} {\bibfnamefont {Y.}~\bibnamefont {Zhu}}, \bibinfo {author} {\bibfnamefont {D.}~\bibnamefont {Peng}}, \bibinfo {author} {\bibfnamefont {E.}~\bibnamefont {Zhang}}, \bibinfo {author} {\bibfnamefont {B.}~\bibnamefont {Pan}}, \bibinfo {author} {\bibfnamefont {X.}~\bibnamefont {Chen}}, \bibinfo {author} {\bibfnamefont {L.}~\bibnamefont {Chen}}, \bibinfo {author} {\bibfnamefont {H.}~\bibnamefont {Ren}}, \bibinfo {author} {\bibfnamefont {F.}~\bibnamefont {Liu}}, \bibinfo {author} {\bibfnamefont {Y.}~\bibnamefont {Hao}}, \bibinfo {author} {\bibfnamefont {N.}~\bibnamefont {Li}}, \bibinfo {author} {\bibfnamefont {Z.}~\bibnamefont {Xing}}, \bibinfo {author} {\bibfnamefont {F.}~\bibnamefont {Lan}}, \bibinfo {author} {\bibfnamefont {J.}~\bibnamefont {Han}}, \bibinfo {author} {\bibfnamefont {J.}~\bibnamefont {Wang}}, \bibinfo {author} {\bibfnamefont {D.}~\bibnamefont {Jia}}, \bibinfo {author} {\bibfnamefont {H.}~\bibnamefont {Wo}}, \bibinfo {author} {\bibfnamefont {Y.}~\bibnamefont
  {Gu}}, \bibinfo {author} {\bibfnamefont {Y.}~\bibnamefont {Gu}}, \bibinfo {author} {\bibfnamefont {L.}~\bibnamefont {Ji}}, \bibinfo {author} {\bibfnamefont {W.}~\bibnamefont {Wang}}, \bibinfo {author} {\bibfnamefont {H.}~\bibnamefont {Gou}}, \bibinfo {author} {\bibfnamefont {Y.}~\bibnamefont {Shen}}, \bibinfo {author} {\bibfnamefont {T.}~\bibnamefont {Ying}}, \bibinfo {author} {\bibfnamefont {X.}~\bibnamefont {Chen}}, \bibinfo {author} {\bibfnamefont {W.}~\bibnamefont {Yang}}, \bibinfo {author} {\bibfnamefont {H.}~\bibnamefont {Cao}}, \bibinfo {author} {\bibfnamefont {C.}~\bibnamefont {Zheng}}, \bibinfo {author} {\bibfnamefont {Q.}~\bibnamefont {Zeng}}, \bibinfo {author} {\bibfnamefont {J.-g.}\ \bibnamefont {Guo}},\ and\ \bibinfo {author} {\bibfnamefont {J.}~\bibnamefont {Zhao}},\ }\href {https://doi.org/10.1038/s41586-024-07553-3} {\bibfield  {journal} {\bibinfo  {journal} {Nature}\ }\textbf {\bibinfo {volume} {631}},\ \bibinfo {pages} {531} (\bibinfo {year} {2024})}\BibitemShut {NoStop}%
\bibitem [{\citenamefont {Sakakibara}\ \emph {et~al.}(2024)\citenamefont {Sakakibara}, \citenamefont {Ochi}, \citenamefont {Nagata}, \citenamefont {Ueki}, \citenamefont {Sakurai}, \citenamefont {Matsumoto}, \citenamefont {Terashima}, \citenamefont {Hirose}, \citenamefont {Ohta}, \citenamefont {Kato}, \citenamefont {Takano},\ and\ \citenamefont {Kuroki}}]{4310-2}%
  \BibitemOpen
  \bibfield  {author} {\bibinfo {author} {\bibfnamefont {H.}~\bibnamefont {Sakakibara}}, \bibinfo {author} {\bibfnamefont {M.}~\bibnamefont {Ochi}}, \bibinfo {author} {\bibfnamefont {H.}~\bibnamefont {Nagata}}, \bibinfo {author} {\bibfnamefont {Y.}~\bibnamefont {Ueki}}, \bibinfo {author} {\bibfnamefont {H.}~\bibnamefont {Sakurai}}, \bibinfo {author} {\bibfnamefont {R.}~\bibnamefont {Matsumoto}}, \bibinfo {author} {\bibfnamefont {K.}~\bibnamefont {Terashima}}, \bibinfo {author} {\bibfnamefont {K.}~\bibnamefont {Hirose}}, \bibinfo {author} {\bibfnamefont {H.}~\bibnamefont {Ohta}}, \bibinfo {author} {\bibfnamefont {M.}~\bibnamefont {Kato}}, \bibinfo {author} {\bibfnamefont {Y.}~\bibnamefont {Takano}},\ and\ \bibinfo {author} {\bibfnamefont {K.}~\bibnamefont {Kuroki}},\ }\href {https://doi.org/10.1103/PhysRevB.109.144511} {\bibfield  {journal} {\bibinfo  {journal} {Phys. Rev. B}\ }\textbf {\bibinfo {volume} {109}},\ \bibinfo {pages} {144511} (\bibinfo {year} {2024})}\BibitemShut {NoStop}%
\bibitem [{\citenamefont {Zhang}\ \emph {et~al.}(2025{\natexlab{a}})\citenamefont {Zhang}, \citenamefont {Pei}, \citenamefont {Peng}, \citenamefont {Du}, \citenamefont {Hu}, \citenamefont {Cao}, \citenamefont {Wang}, \citenamefont {Wu}, \citenamefont {Li}, \citenamefont {Liu}, \citenamefont {Wen}, \citenamefont {Song}, \citenamefont {Zhao}, \citenamefont {Li}, \citenamefont {Cao}, \citenamefont {Zhu}, \citenamefont {Zhang}, \citenamefont {Yu}, \citenamefont {Cheng}, \citenamefont {Zhang}, \citenamefont {Li}, \citenamefont {Zhao}, \citenamefont {Chen}, \citenamefont {Jin}, \citenamefont {Guo}, \citenamefont {Wu}, \citenamefont {Yang}, \citenamefont {Zeng}, \citenamefont {Yan}, \citenamefont {Yang},\ and\ \citenamefont {Qi}}]{4310-3}%
  \BibitemOpen
  \bibfield  {author} {\bibinfo {author} {\bibfnamefont {M.}~\bibnamefont {Zhang}}, \bibinfo {author} {\bibfnamefont {C.}~\bibnamefont {Pei}}, \bibinfo {author} {\bibfnamefont {D.}~\bibnamefont {Peng}}, \bibinfo {author} {\bibfnamefont {X.}~\bibnamefont {Du}}, \bibinfo {author} {\bibfnamefont {W.}~\bibnamefont {Hu}}, \bibinfo {author} {\bibfnamefont {Y.}~\bibnamefont {Cao}}, \bibinfo {author} {\bibfnamefont {Q.}~\bibnamefont {Wang}}, \bibinfo {author} {\bibfnamefont {J.}~\bibnamefont {Wu}}, \bibinfo {author} {\bibfnamefont {Y.}~\bibnamefont {Li}}, \bibinfo {author} {\bibfnamefont {H.}~\bibnamefont {Liu}}, \bibinfo {author} {\bibfnamefont {C.}~\bibnamefont {Wen}}, \bibinfo {author} {\bibfnamefont {J.}~\bibnamefont {Song}}, \bibinfo {author} {\bibfnamefont {Y.}~\bibnamefont {Zhao}}, \bibinfo {author} {\bibfnamefont {C.}~\bibnamefont {Li}}, \bibinfo {author} {\bibfnamefont {W.}~\bibnamefont {Cao}}, \bibinfo {author} {\bibfnamefont {S.}~\bibnamefont {Zhu}}, \bibinfo {author} {\bibfnamefont {Q.}~\bibnamefont
  {Zhang}}, \bibinfo {author} {\bibfnamefont {N.}~\bibnamefont {Yu}}, \bibinfo {author} {\bibfnamefont {P.}~\bibnamefont {Cheng}}, \bibinfo {author} {\bibfnamefont {L.}~\bibnamefont {Zhang}}, \bibinfo {author} {\bibfnamefont {Z.}~\bibnamefont {Li}}, \bibinfo {author} {\bibfnamefont {J.}~\bibnamefont {Zhao}}, \bibinfo {author} {\bibfnamefont {Y.}~\bibnamefont {Chen}}, \bibinfo {author} {\bibfnamefont {C.}~\bibnamefont {Jin}}, \bibinfo {author} {\bibfnamefont {H.}~\bibnamefont {Guo}}, \bibinfo {author} {\bibfnamefont {C.}~\bibnamefont {Wu}}, \bibinfo {author} {\bibfnamefont {F.}~\bibnamefont {Yang}}, \bibinfo {author} {\bibfnamefont {Q.}~\bibnamefont {Zeng}}, \bibinfo {author} {\bibfnamefont {S.}~\bibnamefont {Yan}}, \bibinfo {author} {\bibfnamefont {L.}~\bibnamefont {Yang}},\ and\ \bibinfo {author} {\bibfnamefont {Y.}~\bibnamefont {Qi}},\ }\href {https://doi.org/10.1103/PhysRevX.15.021005} {\bibfield  {journal} {\bibinfo  {journal} {Physical Review X}\ }\textbf {\bibinfo {volume} {15}},\ \bibinfo {pages}
  {021005} (\bibinfo {year} {2025}{\natexlab{a}})}\BibitemShut {NoStop}%
\bibitem [{\citenamefont {Li}\ \emph {et~al.}(2024)\citenamefont {Li}, \citenamefont {Zhang}, \citenamefont {Xiang}, \citenamefont {Zhang}, \citenamefont {Zhu},\ and\ \citenamefont {Wen}}]{4310-4}%
  \BibitemOpen
  \bibfield  {author} {\bibinfo {author} {\bibfnamefont {Q.}~\bibnamefont {Li}}, \bibinfo {author} {\bibfnamefont {Y.-J.}\ \bibnamefont {Zhang}}, \bibinfo {author} {\bibfnamefont {Z.-N.}\ \bibnamefont {Xiang}}, \bibinfo {author} {\bibfnamefont {Y.}~\bibnamefont {Zhang}}, \bibinfo {author} {\bibfnamefont {X.}~\bibnamefont {Zhu}},\ and\ \bibinfo {author} {\bibfnamefont {H.-H.}\ \bibnamefont {Wen}},\ }\href {https://doi.org/10.1088/0256-307X/41/1/017401} {\bibfield  {journal} {\bibinfo  {journal} {Chinese Physics Letters}\ }\textbf {\bibinfo {volume} {41}},\ \bibinfo {pages} {017401} (\bibinfo {year} {2024})}\BibitemShut {NoStop}%
\bibitem [{\citenamefont {Zhang}\ \emph {et~al.}(2025{\natexlab{b}})\citenamefont {Zhang}, \citenamefont {Peng}, \citenamefont {Zhu}, \citenamefont {Chen}, \citenamefont {Cui}, \citenamefont {Wang}, \citenamefont {Wang}, \citenamefont {Zeng},\ and\ \citenamefont {Zhao}}]{4310-5}%
  \BibitemOpen
  \bibfield  {author} {\bibinfo {author} {\bibfnamefont {E.}~\bibnamefont {Zhang}}, \bibinfo {author} {\bibfnamefont {D.}~\bibnamefont {Peng}}, \bibinfo {author} {\bibfnamefont {Y.}~\bibnamefont {Zhu}}, \bibinfo {author} {\bibfnamefont {L.}~\bibnamefont {Chen}}, \bibinfo {author} {\bibfnamefont {B.}~\bibnamefont {Cui}}, \bibinfo {author} {\bibfnamefont {X.}~\bibnamefont {Wang}}, \bibinfo {author} {\bibfnamefont {W.}~\bibnamefont {Wang}}, \bibinfo {author} {\bibfnamefont {Q.}~\bibnamefont {Zeng}},\ and\ \bibinfo {author} {\bibfnamefont {J.}~\bibnamefont {Zhao}},\ }\href {https://doi.org/10.1103/PhysRevX.15.021008} {\bibfield  {journal} {\bibinfo  {journal} {Phys. Rev. X}\ }\textbf {\bibinfo {volume} {15}},\ \bibinfo {pages} {021008} (\bibinfo {year} {2025}{\natexlab{b}})}\BibitemShut {NoStop}%
\bibitem [{\citenamefont {Wang}\ \emph {et~al.}(2025)\citenamefont {Wang}, \citenamefont {Jiang}, \citenamefont {Ying}, \citenamefont {Wu}, \citenamefont {Cheng}, \citenamefont {Hu},\ and\ \citenamefont {Chen}}]{Review}%
  \BibitemOpen
  \bibfield  {author} {\bibinfo {author} {\bibfnamefont {Y.}~\bibnamefont {Wang}}, \bibinfo {author} {\bibfnamefont {K.}~\bibnamefont {Jiang}}, \bibinfo {author} {\bibfnamefont {J.}~\bibnamefont {Ying}}, \bibinfo {author} {\bibfnamefont {T.}~\bibnamefont {Wu}}, \bibinfo {author} {\bibfnamefont {J.}~\bibnamefont {Cheng}}, \bibinfo {author} {\bibfnamefont {J.}~\bibnamefont {Hu}},\ and\ \bibinfo {author} {\bibfnamefont {X.}~\bibnamefont {Chen}},\ }\href {https://doi.org/10.1093/nsr/nwaf373} {\bibfield  {journal} {\bibinfo  {journal} {National Science Review}\ }\textbf {\bibinfo {volume} {12}},\ \bibinfo {pages} {nwaf373} (\bibinfo {year} {2025})}\BibitemShut {NoStop}%
\bibitem [{\citenamefont {Wang}\ \emph {et~al.}(2024{\natexlab{b}})\citenamefont {Wang}, \citenamefont {Wen}, \citenamefont {Wu}, \citenamefont {Yao},\ and\ \citenamefont {Xiang}}]{2024Xiangb}%
  \BibitemOpen
  \bibfield  {author} {\bibinfo {author} {\bibfnamefont {M.}~\bibnamefont {Wang}}, \bibinfo {author} {\bibfnamefont {H.-H.}\ \bibnamefont {Wen}}, \bibinfo {author} {\bibfnamefont {T.}~\bibnamefont {Wu}}, \bibinfo {author} {\bibfnamefont {D.-X.}\ \bibnamefont {Yao}},\ and\ \bibinfo {author} {\bibfnamefont {T.}~\bibnamefont {Xiang}},\ }\href {https://doi.org/10.1088/0256-307X/41/7/077402} {\bibfield  {journal} {\bibinfo  {journal} {Chinese Physics Letters}\ }\textbf {\bibinfo {volume} {41}},\ \bibinfo {pages} {077402} (\bibinfo {year} {2024}{\natexlab{b}})}\BibitemShut {NoStop}%
\bibitem [{\citenamefont {Ko}\ \emph {et~al.}(2025)\citenamefont {Ko}, \citenamefont {Yu}, \citenamefont {Liu}, \citenamefont {Bhatt}, \citenamefont {Li}, \citenamefont {Thampy}, \citenamefont {Kuo}, \citenamefont {Wang}, \citenamefont {Lee}, \citenamefont {Lee}, \citenamefont {Lee}, \citenamefont {Goodge}, \citenamefont {Muller},\ and\ \citenamefont {Hwang}}]{2025Hwangc}%
  \BibitemOpen
  \bibfield  {author} {\bibinfo {author} {\bibfnamefont {E.~K.}\ \bibnamefont {Ko}}, \bibinfo {author} {\bibfnamefont {Y.}~\bibnamefont {Yu}}, \bibinfo {author} {\bibfnamefont {Y.}~\bibnamefont {Liu}}, \bibinfo {author} {\bibfnamefont {L.}~\bibnamefont {Bhatt}}, \bibinfo {author} {\bibfnamefont {J.}~\bibnamefont {Li}}, \bibinfo {author} {\bibfnamefont {V.}~\bibnamefont {Thampy}}, \bibinfo {author} {\bibfnamefont {C.-T.}\ \bibnamefont {Kuo}}, \bibinfo {author} {\bibfnamefont {B.~Y.}\ \bibnamefont {Wang}}, \bibinfo {author} {\bibfnamefont {Y.}~\bibnamefont {Lee}}, \bibinfo {author} {\bibfnamefont {K.}~\bibnamefont {Lee}}, \bibinfo {author} {\bibfnamefont {J.-S.}\ \bibnamefont {Lee}}, \bibinfo {author} {\bibfnamefont {B.~H.}\ \bibnamefont {Goodge}}, \bibinfo {author} {\bibfnamefont {D.~A.}\ \bibnamefont {Muller}},\ and\ \bibinfo {author} {\bibfnamefont {H.~Y.}\ \bibnamefont {Hwang}},\ }\href {https://doi.org/10.1038/s41586-024-08525-3} {\bibfield  {journal} {\bibinfo  {journal} {Nature}\ }\textbf {\bibinfo
  {volume} {638}},\ \bibinfo {pages} {935} (\bibinfo {year} {2025})}\BibitemShut {NoStop}%
\bibitem [{\citenamefont {Liu}\ \emph {et~al.}(2025{\natexlab{a}})\citenamefont {Liu}, \citenamefont {Ko}, \citenamefont {Tarn}, \citenamefont {Bhatt}, \citenamefont {Li}, \citenamefont {Thampy}, \citenamefont {Goodge}, \citenamefont {Muller}, \citenamefont {Raghu}, \citenamefont {Yu},\ and\ \citenamefont {Hwang}}]{2025Hwang}%
  \BibitemOpen
  \bibfield  {author} {\bibinfo {author} {\bibfnamefont {Y.}~\bibnamefont {Liu}}, \bibinfo {author} {\bibfnamefont {E.~K.}\ \bibnamefont {Ko}}, \bibinfo {author} {\bibfnamefont {Y.}~\bibnamefont {Tarn}}, \bibinfo {author} {\bibfnamefont {L.}~\bibnamefont {Bhatt}}, \bibinfo {author} {\bibfnamefont {J.}~\bibnamefont {Li}}, \bibinfo {author} {\bibfnamefont {V.}~\bibnamefont {Thampy}}, \bibinfo {author} {\bibfnamefont {B.~H.}\ \bibnamefont {Goodge}}, \bibinfo {author} {\bibfnamefont {D.~A.}\ \bibnamefont {Muller}}, \bibinfo {author} {\bibfnamefont {S.}~\bibnamefont {Raghu}}, \bibinfo {author} {\bibfnamefont {Y.}~\bibnamefont {Yu}},\ and\ \bibinfo {author} {\bibfnamefont {H.~Y.}\ \bibnamefont {Hwang}},\ }\href {https://doi.org/10.1038/s41563-025-02258-y} {\bibfield  {journal} {\bibinfo  {journal} {Nature Materials}\ }\textbf {\bibinfo {volume} {24}},\ \bibinfo {pages} {1221} (\bibinfo {year} {2025}{\natexlab{a}})}\BibitemShut {NoStop}%
\bibitem [{\citenamefont {Zhou}\ \emph {et~al.}(2025)\citenamefont {Zhou}, \citenamefont {Lv}, \citenamefont {Wang}, \citenamefont {Nie}, \citenamefont {Chen}, \citenamefont {Li}, \citenamefont {Huang}, \citenamefont {Chen}, \citenamefont {Sun}, \citenamefont {Xue},\ and\ \citenamefont {Chen}}]{2025Chenb}%
  \BibitemOpen
  \bibfield  {author} {\bibinfo {author} {\bibfnamefont {G.}~\bibnamefont {Zhou}}, \bibinfo {author} {\bibfnamefont {W.}~\bibnamefont {Lv}}, \bibinfo {author} {\bibfnamefont {H.}~\bibnamefont {Wang}}, \bibinfo {author} {\bibfnamefont {Z.}~\bibnamefont {Nie}}, \bibinfo {author} {\bibfnamefont {Y.}~\bibnamefont {Chen}}, \bibinfo {author} {\bibfnamefont {Y.}~\bibnamefont {Li}}, \bibinfo {author} {\bibfnamefont {H.}~\bibnamefont {Huang}}, \bibinfo {author} {\bibfnamefont {W.-Q.}\ \bibnamefont {Chen}}, \bibinfo {author} {\bibfnamefont {Y.-J.}\ \bibnamefont {Sun}}, \bibinfo {author} {\bibfnamefont {Q.-K.}\ \bibnamefont {Xue}},\ and\ \bibinfo {author} {\bibfnamefont {Z.}~\bibnamefont {Chen}},\ }\href {https://doi.org/10.1038/s41586-025-08755-z} {\bibfield  {journal} {\bibinfo  {journal} {Nature}\ }\textbf {\bibinfo {volume} {640}},\ \bibinfo {pages} {641} (\bibinfo {year} {2025})}\BibitemShut {NoStop}%
\bibitem [{\citenamefont {Shi}\ \emph {et~al.}(2025)\citenamefont {Shi}, \citenamefont {Peng}, \citenamefont {Fan}, \citenamefont {Xing}, \citenamefont {Yang}, \citenamefont {Wang}, \citenamefont {Li}, \citenamefont {Wu}, \citenamefont {Du}, \citenamefont {Ge}, \citenamefont {Zeng}, \citenamefont {Zeng}, \citenamefont {Ying}, \citenamefont {Wu},\ and\ \citenamefont {Chen}}]{2025Chenc}%
  \BibitemOpen
  \bibfield  {author} {\bibinfo {author} {\bibfnamefont {M.}~\bibnamefont {Shi}}, \bibinfo {author} {\bibfnamefont {D.}~\bibnamefont {Peng}}, \bibinfo {author} {\bibfnamefont {K.}~\bibnamefont {Fan}}, \bibinfo {author} {\bibfnamefont {Z.}~\bibnamefont {Xing}}, \bibinfo {author} {\bibfnamefont {S.}~\bibnamefont {Yang}}, \bibinfo {author} {\bibfnamefont {Y.}~\bibnamefont {Wang}}, \bibinfo {author} {\bibfnamefont {H.}~\bibnamefont {Li}}, \bibinfo {author} {\bibfnamefont {R.}~\bibnamefont {Wu}}, \bibinfo {author} {\bibfnamefont {M.}~\bibnamefont {Du}}, \bibinfo {author} {\bibfnamefont {B.}~\bibnamefont {Ge}}, \bibinfo {author} {\bibfnamefont {Z.}~\bibnamefont {Zeng}}, \bibinfo {author} {\bibfnamefont {Q.}~\bibnamefont {Zeng}}, \bibinfo {author} {\bibfnamefont {J.}~\bibnamefont {Ying}}, \bibinfo {author} {\bibfnamefont {T.}~\bibnamefont {Wu}},\ and\ \bibinfo {author} {\bibfnamefont {X.}~\bibnamefont {Chen}},\ }\href {https://doi.org/10.1038/s41567-025-03023-3} {\bibfield  {journal} {\bibinfo  {journal} {Nature
  Physics}\ }\textbf {\bibinfo {volume} {21}},\ \bibinfo {pages} {1780} (\bibinfo {year} {2025})}\BibitemShut {NoStop}%
\bibitem [{\citenamefont {Liu}\ \emph {et~al.}(2022)\citenamefont {Liu}, \citenamefont {Sun}, \citenamefont {Huo}, \citenamefont {Ma}, \citenamefont {Ji}, \citenamefont {Yi}, \citenamefont {Li}, \citenamefont {Liu}, \citenamefont {Yu}, \citenamefont {Zhang}, \citenamefont {Chen}, \citenamefont {Liang}, \citenamefont {Dong}, \citenamefont {Guo}, \citenamefont {Zhong}, \citenamefont {Shen}, \citenamefont {Li},\ and\ \citenamefont {Wang}}]{wangmeng_sdw}%
  \BibitemOpen
  \bibfield  {author} {\bibinfo {author} {\bibfnamefont {Z.}~\bibnamefont {Liu}}, \bibinfo {author} {\bibfnamefont {H.}~\bibnamefont {Sun}}, \bibinfo {author} {\bibfnamefont {M.}~\bibnamefont {Huo}}, \bibinfo {author} {\bibfnamefont {X.}~\bibnamefont {Ma}}, \bibinfo {author} {\bibfnamefont {Y.}~\bibnamefont {Ji}}, \bibinfo {author} {\bibfnamefont {E.}~\bibnamefont {Yi}}, \bibinfo {author} {\bibfnamefont {L.}~\bibnamefont {Li}}, \bibinfo {author} {\bibfnamefont {H.}~\bibnamefont {Liu}}, \bibinfo {author} {\bibfnamefont {J.}~\bibnamefont {Yu}}, \bibinfo {author} {\bibfnamefont {Z.}~\bibnamefont {Zhang}}, \bibinfo {author} {\bibfnamefont {Z.}~\bibnamefont {Chen}}, \bibinfo {author} {\bibfnamefont {F.}~\bibnamefont {Liang}}, \bibinfo {author} {\bibfnamefont {H.}~\bibnamefont {Dong}}, \bibinfo {author} {\bibfnamefont {H.}~\bibnamefont {Guo}}, \bibinfo {author} {\bibfnamefont {D.}~\bibnamefont {Zhong}}, \bibinfo {author} {\bibfnamefont {B.}~\bibnamefont {Shen}}, \bibinfo {author} {\bibfnamefont {S.}~\bibnamefont
  {Li}},\ and\ \bibinfo {author} {\bibfnamefont {M.}~\bibnamefont {Wang}},\ }\href {https://doi.org/10.1007/s11433-022-1962-4} {\bibfield  {journal} {\bibinfo  {journal} {Science China Physics, Mechanics \& Astronomy}\ }\textbf {\bibinfo {volume} {66}},\ \bibinfo {pages} {217411} (\bibinfo {year} {2022})}\BibitemShut {NoStop}%
\bibitem [{\citenamefont {Zhao}\ \emph {et~al.}(2025)\citenamefont {Zhao}, \citenamefont {Zhou}, \citenamefont {Huo}, \citenamefont {Wang}, \citenamefont {Nie}, \citenamefont {Yang}, \citenamefont {Ying}, \citenamefont {Wang}, \citenamefont {Wu},\ and\ \citenamefont {Chen}}]{taowu_327}%
  \BibitemOpen
  \bibfield  {author} {\bibinfo {author} {\bibfnamefont {D.}~\bibnamefont {Zhao}}, \bibinfo {author} {\bibfnamefont {Y.}~\bibnamefont {Zhou}}, \bibinfo {author} {\bibfnamefont {M.}~\bibnamefont {Huo}}, \bibinfo {author} {\bibfnamefont {Y.}~\bibnamefont {Wang}}, \bibinfo {author} {\bibfnamefont {L.}~\bibnamefont {Nie}}, \bibinfo {author} {\bibfnamefont {Y.}~\bibnamefont {Yang}}, \bibinfo {author} {\bibfnamefont {J.}~\bibnamefont {Ying}}, \bibinfo {author} {\bibfnamefont {M.}~\bibnamefont {Wang}}, \bibinfo {author} {\bibfnamefont {T.}~\bibnamefont {Wu}},\ and\ \bibinfo {author} {\bibfnamefont {X.}~\bibnamefont {Chen}},\ }\href {https://doi.org/https://doi.org/10.1016/j.scib.2025.02.019} {\bibfield  {journal} {\bibinfo  {journal} {Science Bulletin}\ }\textbf {\bibinfo {volume} {70}},\ \bibinfo {pages} {1239} (\bibinfo {year} {2025})}\BibitemShut {NoStop}%
\bibitem [{\citenamefont {Chen}\ \emph {et~al.}(2024{\natexlab{a}})\citenamefont {Chen}, \citenamefont {Liu}, \citenamefont {Jiao}, \citenamefont {Zou}, \citenamefont {Jiang}, \citenamefont {Li}, \citenamefont {Luo}, \citenamefont {Wu}, \citenamefont {Zhang}, \citenamefont {Guo},\ and\ \citenamefont {Shu}}]{shulei_PhysRevLett.132.256503}%
  \BibitemOpen
  \bibfield  {author} {\bibinfo {author} {\bibfnamefont {K.}~\bibnamefont {Chen}}, \bibinfo {author} {\bibfnamefont {X.}~\bibnamefont {Liu}}, \bibinfo {author} {\bibfnamefont {J.}~\bibnamefont {Jiao}}, \bibinfo {author} {\bibfnamefont {M.}~\bibnamefont {Zou}}, \bibinfo {author} {\bibfnamefont {C.}~\bibnamefont {Jiang}}, \bibinfo {author} {\bibfnamefont {X.}~\bibnamefont {Li}}, \bibinfo {author} {\bibfnamefont {Y.}~\bibnamefont {Luo}}, \bibinfo {author} {\bibfnamefont {Q.}~\bibnamefont {Wu}}, \bibinfo {author} {\bibfnamefont {N.}~\bibnamefont {Zhang}}, \bibinfo {author} {\bibfnamefont {Y.}~\bibnamefont {Guo}},\ and\ \bibinfo {author} {\bibfnamefont {L.}~\bibnamefont {Shu}},\ }\href {https://doi.org/10.1103/PhysRevLett.132.256503} {\bibfield  {journal} {\bibinfo  {journal} {Phys. Rev. Lett.}\ }\textbf {\bibinfo {volume} {132}},\ \bibinfo {pages} {256503} (\bibinfo {year} {2024}{\natexlab{a}})}\BibitemShut {NoStop}%
\bibitem [{\citenamefont {Chen}\ \emph {et~al.}(2024{\natexlab{b}})\citenamefont {Chen}, \citenamefont {Choi}, \citenamefont {Jiang}, \citenamefont {Mei}, \citenamefont {Jiang}, \citenamefont {Li}, \citenamefont {Agrestini}, \citenamefont {Garcia-Fernandez}, \citenamefont {Sun}, \citenamefont {Huang}, \citenamefont {Shen}, \citenamefont {Wang}, \citenamefont {Hu}, \citenamefont {Lu}, \citenamefont {Zhou},\ and\ \citenamefont {Feng}}]{Chen2024}%
  \BibitemOpen
  \bibfield  {author} {\bibinfo {author} {\bibfnamefont {X.}~\bibnamefont {Chen}}, \bibinfo {author} {\bibfnamefont {J.}~\bibnamefont {Choi}}, \bibinfo {author} {\bibfnamefont {Z.}~\bibnamefont {Jiang}}, \bibinfo {author} {\bibfnamefont {J.}~\bibnamefont {Mei}}, \bibinfo {author} {\bibfnamefont {K.}~\bibnamefont {Jiang}}, \bibinfo {author} {\bibfnamefont {J.}~\bibnamefont {Li}}, \bibinfo {author} {\bibfnamefont {S.}~\bibnamefont {Agrestini}}, \bibinfo {author} {\bibfnamefont {M.}~\bibnamefont {Garcia-Fernandez}}, \bibinfo {author} {\bibfnamefont {H.}~\bibnamefont {Sun}}, \bibinfo {author} {\bibfnamefont {X.}~\bibnamefont {Huang}}, \bibinfo {author} {\bibfnamefont {D.}~\bibnamefont {Shen}}, \bibinfo {author} {\bibfnamefont {M.}~\bibnamefont {Wang}}, \bibinfo {author} {\bibfnamefont {J.}~\bibnamefont {Hu}}, \bibinfo {author} {\bibfnamefont {Y.}~\bibnamefont {Lu}}, \bibinfo {author} {\bibfnamefont {K.-J.}\ \bibnamefont {Zhou}},\ and\ \bibinfo {author} {\bibfnamefont {D.}~\bibnamefont {Feng}},\ }\href
  {https://doi.org/10.1038/s41467-024-53863-5} {\bibfield  {journal} {\bibinfo  {journal} {Nature Communications}\ }\textbf {\bibinfo {volume} {15}},\ \bibinfo {pages} {9597} (\bibinfo {year} {2024}{\natexlab{b}})}\BibitemShut {NoStop}%
\bibitem [{\citenamefont {Khasanov}\ \emph {et~al.}(2025)\citenamefont {Khasanov}, \citenamefont {Hicken}, \citenamefont {Gawryluk}, \citenamefont {Sazgari}, \citenamefont {Plokhikh}, \citenamefont {Sorel}, \citenamefont {Bartkowiak}, \citenamefont {B{\"o}tzel}, \citenamefont {Lechermann}, \citenamefont {Eremin}, \citenamefont {Luetkens},\ and\ \citenamefont {Guguchia}}]{Khasanov2025}%
  \BibitemOpen
  \bibfield  {author} {\bibinfo {author} {\bibfnamefont {R.}~\bibnamefont {Khasanov}}, \bibinfo {author} {\bibfnamefont {T.~J.}\ \bibnamefont {Hicken}}, \bibinfo {author} {\bibfnamefont {D.~J.}\ \bibnamefont {Gawryluk}}, \bibinfo {author} {\bibfnamefont {V.}~\bibnamefont {Sazgari}}, \bibinfo {author} {\bibfnamefont {I.}~\bibnamefont {Plokhikh}}, \bibinfo {author} {\bibfnamefont {L.~P.}\ \bibnamefont {Sorel}}, \bibinfo {author} {\bibfnamefont {M.}~\bibnamefont {Bartkowiak}}, \bibinfo {author} {\bibfnamefont {S.}~\bibnamefont {B{\"o}tzel}}, \bibinfo {author} {\bibfnamefont {F.}~\bibnamefont {Lechermann}}, \bibinfo {author} {\bibfnamefont {I.~M.}\ \bibnamefont {Eremin}}, \bibinfo {author} {\bibfnamefont {H.}~\bibnamefont {Luetkens}},\ and\ \bibinfo {author} {\bibfnamefont {Z.}~\bibnamefont {Guguchia}},\ }\href {https://doi.org/10.1038/s41567-024-02754-z} {\bibfield  {journal} {\bibinfo  {journal} {Nature Physics}\ }\textbf {\bibinfo {volume} {21}},\ \bibinfo {pages} {430} (\bibinfo {year} {2025})}\BibitemShut
  {NoStop}%
\bibitem [{\citenamefont {Wang}\ \emph {et~al.}(2026)\citenamefont {Wang}, \citenamefont {Zhao}, \citenamefont {Zhang}, \citenamefont {Chen}, \citenamefont {Zhou}, \citenamefont {Shi}, \citenamefont {Zhu}, \citenamefont {Ying}, \citenamefont {Zhao},\ and\ \citenamefont {Wu}}]{taowu-4310}%
  \BibitemOpen
  \bibfield  {author} {\bibinfo {author} {\bibfnamefont {Y.}~\bibnamefont {Wang}}, \bibinfo {author} {\bibfnamefont {D.}~\bibnamefont {Zhao}}, \bibinfo {author} {\bibfnamefont {E.}~\bibnamefont {Zhang}}, \bibinfo {author} {\bibfnamefont {L.}~\bibnamefont {Chen}}, \bibinfo {author} {\bibfnamefont {Y.}~\bibnamefont {Zhou}}, \bibinfo {author} {\bibfnamefont {M.}~\bibnamefont {Shi}}, \bibinfo {author} {\bibfnamefont {Y.}~\bibnamefont {Zhu}}, \bibinfo {author} {\bibfnamefont {J.}~\bibnamefont {Ying}}, \bibinfo {author} {\bibfnamefont {J.}~\bibnamefont {Zhao}},\ and\ \bibinfo {author} {\bibfnamefont {T.}~\bibnamefont {Wu}},\ }\bibfield  {journal} {\bibinfo  {journal} {Nature Communications}\ }\href {https://doi.org/10.1038/s41467-026-73082-4} {10.1038/s41467-026-73082-4} (\bibinfo {year} {2026})\BibitemShut {NoStop}%
\bibitem [{\citenamefont {Yashima}\ \emph {et~al.}(2025)\citenamefont {Yashima}, \citenamefont {Seto}, \citenamefont {Oshita}, \citenamefont {Kakoi}, \citenamefont {Sakurai}, \citenamefont {Takano},\ and\ \citenamefont {Mukuda}}]{JPSJ.94.054704}%
  \BibitemOpen
  \bibfield  {author} {\bibinfo {author} {\bibfnamefont {M.}~\bibnamefont {Yashima}}, \bibinfo {author} {\bibfnamefont {N.}~\bibnamefont {Seto}}, \bibinfo {author} {\bibfnamefont {Y.}~\bibnamefont {Oshita}}, \bibinfo {author} {\bibfnamefont {M.}~\bibnamefont {Kakoi}}, \bibinfo {author} {\bibfnamefont {H.}~\bibnamefont {Sakurai}}, \bibinfo {author} {\bibfnamefont {Y.}~\bibnamefont {Takano}},\ and\ \bibinfo {author} {\bibfnamefont {H.}~\bibnamefont {Mukuda}},\ }\href {https://doi.org/10.7566/JPSJ.94.054704} {\bibfield  {journal} {\bibinfo  {journal} {Journal of the Physical Society of Japan}\ }\textbf {\bibinfo {volume} {94}},\ \bibinfo {pages} {054704} (\bibinfo {year} {2025})}\BibitemShut {NoStop}%
\bibitem [{\citenamefont {Ren}\ \emph {et~al.}(2025)\citenamefont {Ren}, \citenamefont {Sutarto}, \citenamefont {Wu}, \citenamefont {Zhang}, \citenamefont {Huang}, \citenamefont {Xiang}, \citenamefont {Hu}, \citenamefont {Comin}, \citenamefont {Zhou},\ and\ \citenamefont {Zhu}}]{RenXL2025}%
  \BibitemOpen
  \bibfield  {author} {\bibinfo {author} {\bibfnamefont {X.}~\bibnamefont {Ren}}, \bibinfo {author} {\bibfnamefont {R.}~\bibnamefont {Sutarto}}, \bibinfo {author} {\bibfnamefont {X.}~\bibnamefont {Wu}}, \bibinfo {author} {\bibfnamefont {J.}~\bibnamefont {Zhang}}, \bibinfo {author} {\bibfnamefont {H.}~\bibnamefont {Huang}}, \bibinfo {author} {\bibfnamefont {T.}~\bibnamefont {Xiang}}, \bibinfo {author} {\bibfnamefont {J.}~\bibnamefont {Hu}}, \bibinfo {author} {\bibfnamefont {R.}~\bibnamefont {Comin}}, \bibinfo {author} {\bibfnamefont {X.}~\bibnamefont {Zhou}},\ and\ \bibinfo {author} {\bibfnamefont {Z.}~\bibnamefont {Zhu}},\ }\href {https://doi.org/10.1038/s42005-025-01971-z} {\bibfield  {journal} {\bibinfo  {journal} {Communications Physics}\ }\textbf {\bibinfo {volume} {8}},\ \bibinfo {pages} {52} (\bibinfo {year} {2025})}\BibitemShut {NoStop}%
\bibitem [{\citenamefont {Luo}\ \emph {et~al.}(2025)\citenamefont {Luo}, \citenamefont {Feng}, \citenamefont {Wang}, \citenamefont {Wang}, \citenamefont {Dou}, \citenamefont {Fang}, \citenamefont {Yang}, \citenamefont {Cheng}, \citenamefont {Zheng},\ and\ \citenamefont {Zhou}}]{Luo_2025}%
  \BibitemOpen
  \bibfield  {author} {\bibinfo {author} {\bibfnamefont {J.}~\bibnamefont {Luo}}, \bibinfo {author} {\bibfnamefont {J.}~\bibnamefont {Feng}}, \bibinfo {author} {\bibfnamefont {G.}~\bibnamefont {Wang}}, \bibinfo {author} {\bibfnamefont {N.}~\bibnamefont {Wang}}, \bibinfo {author} {\bibfnamefont {J.}~\bibnamefont {Dou}}, \bibinfo {author} {\bibfnamefont {A.}~\bibnamefont {Fang}}, \bibinfo {author} {\bibfnamefont {J.}~\bibnamefont {Yang}}, \bibinfo {author} {\bibfnamefont {J.}~\bibnamefont {Cheng}}, \bibinfo {author} {\bibfnamefont {G.}~\bibnamefont {Zheng}},\ and\ \bibinfo {author} {\bibfnamefont {R.}~\bibnamefont {Zhou}},\ }\href {https://doi.org/10.1088/0256-307X/42/6/067402} {\bibfield  {journal} {\bibinfo  {journal} {Chinese Physics Letters}\ }\textbf {\bibinfo {volume} {42}},\ \bibinfo {pages} {067402} (\bibinfo {year} {2025})}\BibitemShut {NoStop}%
\bibitem [{\citenamefont {Chen}\ \emph {et~al.}(2026{\natexlab{a}})\citenamefont {Chen}, \citenamefont {Zhang}, \citenamefont {Hao}, \citenamefont {Zhu}, \citenamefont {Cui}, \citenamefont {Abernathy}, \citenamefont {Williams}, \citenamefont {Ikeda}, \citenamefont {Zhang}, \citenamefont {Liu}, \citenamefont {Wang}, \citenamefont {Wang},\ and\ \citenamefont {Zhao}}]{chen2026naturemagnetismbilayernickelate}%
  \BibitemOpen
  \bibfield  {author} {\bibinfo {author} {\bibfnamefont {L.}~\bibnamefont {Chen}}, \bibinfo {author} {\bibfnamefont {E.}~\bibnamefont {Zhang}}, \bibinfo {author} {\bibfnamefont {Y.}~\bibnamefont {Hao}}, \bibinfo {author} {\bibfnamefont {Y.}~\bibnamefont {Zhu}}, \bibinfo {author} {\bibfnamefont {B.}~\bibnamefont {Cui}}, \bibinfo {author} {\bibfnamefont {D.~L.}\ \bibnamefont {Abernathy}}, \bibinfo {author} {\bibfnamefont {T.~J.}\ \bibnamefont {Williams}}, \bibinfo {author} {\bibfnamefont {Y.}~\bibnamefont {Ikeda}}, \bibinfo {author} {\bibfnamefont {H.}~\bibnamefont {Zhang}}, \bibinfo {author} {\bibfnamefont {F.}~\bibnamefont {Liu}}, \bibinfo {author} {\bibfnamefont {W.}~\bibnamefont {Wang}}, \bibinfo {author} {\bibfnamefont {Q.}~\bibnamefont {Wang}},\ and\ \bibinfo {author} {\bibfnamefont {J.}~\bibnamefont {Zhao}},\ }\href {https://arxiv.org/abs/2605.03448} {\bibinfo {title} {Nature of magnetism in bilayer nickelate {La$_3$Ni$_2$O$_7$} single crystals}} (\bibinfo {year} {2026}{\natexlab{a}}),\ \Eprint
  {https://arxiv.org/abs/2605.03448} {arXiv:2605.03448 [cond-mat.str-el]} \BibitemShut {NoStop}%
\bibitem [{\citenamefont {Chen}\ \emph {et~al.}(2026{\natexlab{b}})\citenamefont {Chen}, \citenamefont {Li}, \citenamefont {Xie}, \citenamefont {Hu}, \citenamefont {Chiu}, \citenamefont {Agrestini}, \citenamefont {Zhang}, \citenamefont {Lu}, \citenamefont {Wang}, \citenamefont {Garcia-Fernandez}, \citenamefont {Feng},\ and\ \citenamefont {Zhou}}]{chen2026_4310}%
  \BibitemOpen
  \bibfield  {author} {\bibinfo {author} {\bibfnamefont {X.}~\bibnamefont {Chen}}, \bibinfo {author} {\bibfnamefont {Z.}~\bibnamefont {Li}}, \bibinfo {author} {\bibfnamefont {M.}~\bibnamefont {Xie}}, \bibinfo {author} {\bibfnamefont {D.}~\bibnamefont {Hu}}, \bibinfo {author} {\bibfnamefont {Y.-F.}\ \bibnamefont {Chiu}}, \bibinfo {author} {\bibfnamefont {S.}~\bibnamefont {Agrestini}}, \bibinfo {author} {\bibfnamefont {W.}~\bibnamefont {Zhang}}, \bibinfo {author} {\bibfnamefont {Y.}~\bibnamefont {Lu}}, \bibinfo {author} {\bibfnamefont {M.}~\bibnamefont {Wang}}, \bibinfo {author} {\bibfnamefont {M.}~\bibnamefont {Garcia-Fernandez}}, \bibinfo {author} {\bibfnamefont {D.}~\bibnamefont {Feng}},\ and\ \bibinfo {author} {\bibfnamefont {K.-J.}\ \bibnamefont {Zhou}},\ }\href {https://arxiv.org/abs/2604.01902} {\bibinfo {title} {Dissecting superconductivity in the ruddlesden-popper nickelates: The role of electron correlation and interlayer magnetic exchange}} (\bibinfo {year} {2026}{\natexlab{b}}),\ \Eprint
  {https://arxiv.org/abs/2604.01902} {arXiv:2604.01902 [cond-mat.supr-con]} \BibitemShut {NoStop}%
\bibitem [{\citenamefont {Chan}\ \emph {et~al.}(2026)\citenamefont {Chan}, \citenamefont {Li}, \citenamefont {Yan}, \citenamefont {Hong}, \citenamefont {Wang}, \citenamefont {dos Reis~Cantarino}, \citenamefont {Zhu}, \citenamefont {Zhang}, \citenamefont {Chen}, \citenamefont {Okamoto}, \citenamefont {Huang}, \citenamefont {Huang}, \citenamefont {Brookes}, \citenamefont {Chang}, \citenamefont {Shen}, \citenamefont {Zhao},\ and\ \citenamefont {Wang}}]{chan2026collectivespinexcitationstrilayer}%
  \BibitemOpen
  \bibfield  {author} {\bibinfo {author} {\bibfnamefont {Y.}~\bibnamefont {Chan}}, \bibinfo {author} {\bibfnamefont {Y.}~\bibnamefont {Li}}, \bibinfo {author} {\bibfnamefont {Y.}~\bibnamefont {Yan}}, \bibinfo {author} {\bibfnamefont {X.}~\bibnamefont {Hong}}, \bibinfo {author} {\bibfnamefont {T.}~\bibnamefont {Wang}}, \bibinfo {author} {\bibfnamefont {M.}~\bibnamefont {dos Reis~Cantarino}}, \bibinfo {author} {\bibfnamefont {Y.}~\bibnamefont {Zhu}}, \bibinfo {author} {\bibfnamefont {E.}~\bibnamefont {Zhang}}, \bibinfo {author} {\bibfnamefont {L.}~\bibnamefont {Chen}}, \bibinfo {author} {\bibfnamefont {J.}~\bibnamefont {Okamoto}}, \bibinfo {author} {\bibfnamefont {H.-Y.}\ \bibnamefont {Huang}}, \bibinfo {author} {\bibfnamefont {D.-J.}\ \bibnamefont {Huang}}, \bibinfo {author} {\bibfnamefont {N.~B.}\ \bibnamefont {Brookes}}, \bibinfo {author} {\bibfnamefont {J.}~\bibnamefont {Chang}}, \bibinfo {author} {\bibfnamefont {Y.}~\bibnamefont {Shen}}, \bibinfo {author} {\bibfnamefont {J.}~\bibnamefont {Zhao}},\ and\
  \bibinfo {author} {\bibfnamefont {Q.}~\bibnamefont {Wang}},\ }\href {https://arxiv.org/abs/2604.04643} {\bibinfo {title} {Collective spin excitations in trilayer nickelate {La$_4$Ni$_3$O$_{10}$}}} (\bibinfo {year} {2026}),\ \Eprint {https://arxiv.org/abs/2604.04643} {arXiv:2604.04643 [cond-mat.supr-con]} \BibitemShut {NoStop}%
\bibitem [{\citenamefont {Liu}\ \emph {et~al.}(2025{\natexlab{b}})\citenamefont {Liu}, \citenamefont {Sun}, \citenamefont {Zhang}, \citenamefont {Liu}, \citenamefont {Chen},\ and\ \citenamefont {Yang}}]{24f4-349n}%
  \BibitemOpen
  \bibfield  {author} {\bibinfo {author} {\bibfnamefont {Y.-B.}\ \bibnamefont {Liu}}, \bibinfo {author} {\bibfnamefont {H.}~\bibnamefont {Sun}}, \bibinfo {author} {\bibfnamefont {M.}~\bibnamefont {Zhang}}, \bibinfo {author} {\bibfnamefont {Q.}~\bibnamefont {Liu}}, \bibinfo {author} {\bibfnamefont {W.-Q.}\ \bibnamefont {Chen}},\ and\ \bibinfo {author} {\bibfnamefont {F.}~\bibnamefont {Yang}},\ }\href {https://doi.org/10.1103/24f4-349n} {\bibfield  {journal} {\bibinfo  {journal} {Phys. Rev. B}\ }\textbf {\bibinfo {volume} {112}},\ \bibinfo {pages} {014510} (\bibinfo {year} {2025}{\natexlab{b}})}\BibitemShut {NoStop}%
\bibitem [{\citenamefont {Zhang}\ \emph {et~al.}(2025{\natexlab{c}})\citenamefont {Zhang}, \citenamefont {Sun}, \citenamefont {Liu}, \citenamefont {Liu}, \citenamefont {Chen},\ and\ \citenamefont {Yang}}]{PhysRevB.111.144502}%
  \BibitemOpen
  \bibfield  {author} {\bibinfo {author} {\bibfnamefont {M.}~\bibnamefont {Zhang}}, \bibinfo {author} {\bibfnamefont {H.}~\bibnamefont {Sun}}, \bibinfo {author} {\bibfnamefont {Y.-B.}\ \bibnamefont {Liu}}, \bibinfo {author} {\bibfnamefont {Q.}~\bibnamefont {Liu}}, \bibinfo {author} {\bibfnamefont {W.-Q.}\ \bibnamefont {Chen}},\ and\ \bibinfo {author} {\bibfnamefont {F.}~\bibnamefont {Yang}},\ }\href {https://doi.org/10.1103/PhysRevB.111.144502} {\bibfield  {journal} {\bibinfo  {journal} {Phys. Rev. B}\ }\textbf {\bibinfo {volume} {111}},\ \bibinfo {pages} {144502} (\bibinfo {year} {2025}{\natexlab{c}})}\BibitemShut {NoStop}%
\bibitem [{\citenamefont {B\"otzel}\ \emph {et~al.}(2024)\citenamefont {B\"otzel}, \citenamefont {Lechermann}, \citenamefont {Gondolf},\ and\ \citenamefont {Eremin}}]{PhysRevB.109.L180502}%
  \BibitemOpen
  \bibfield  {author} {\bibinfo {author} {\bibfnamefont {S.}~\bibnamefont {B\"otzel}}, \bibinfo {author} {\bibfnamefont {F.}~\bibnamefont {Lechermann}}, \bibinfo {author} {\bibfnamefont {J.}~\bibnamefont {Gondolf}},\ and\ \bibinfo {author} {\bibfnamefont {I.~M.}\ \bibnamefont {Eremin}},\ }\href {https://doi.org/10.1103/PhysRevB.109.L180502} {\bibfield  {journal} {\bibinfo  {journal} {Phys. Rev. B}\ }\textbf {\bibinfo {volume} {109}},\ \bibinfo {pages} {L180502} (\bibinfo {year} {2024})}\BibitemShut {NoStop}%
\bibitem [{\citenamefont {Zhang}\ \emph {et~al.}(2020)\citenamefont {Zhang}, \citenamefont {Phelan}, \citenamefont {Botana}, \citenamefont {Chen}, \citenamefont {Zheng}, \citenamefont {Krogstad}, \citenamefont {Wang}, \citenamefont {Qiu}, \citenamefont {Rodriguez-Rivera}, \citenamefont {Osborn}, \citenamefont {Rosenkranz}, \citenamefont {Norman},\ and\ \citenamefont {Mitchell}}]{Zhang2020}%
  \BibitemOpen
  \bibfield  {author} {\bibinfo {author} {\bibfnamefont {J.}~\bibnamefont {Zhang}}, \bibinfo {author} {\bibfnamefont {D.}~\bibnamefont {Phelan}}, \bibinfo {author} {\bibfnamefont {A.~S.}\ \bibnamefont {Botana}}, \bibinfo {author} {\bibfnamefont {Y.-S.}\ \bibnamefont {Chen}}, \bibinfo {author} {\bibfnamefont {H.}~\bibnamefont {Zheng}}, \bibinfo {author} {\bibfnamefont {M.}~\bibnamefont {Krogstad}}, \bibinfo {author} {\bibfnamefont {S.~G.}\ \bibnamefont {Wang}}, \bibinfo {author} {\bibfnamefont {Y.}~\bibnamefont {Qiu}}, \bibinfo {author} {\bibfnamefont {J.~A.}\ \bibnamefont {Rodriguez-Rivera}}, \bibinfo {author} {\bibfnamefont {R.}~\bibnamefont {Osborn}}, \bibinfo {author} {\bibfnamefont {S.}~\bibnamefont {Rosenkranz}}, \bibinfo {author} {\bibfnamefont {M.~R.}\ \bibnamefont {Norman}},\ and\ \bibinfo {author} {\bibfnamefont {J.~F.}\ \bibnamefont {Mitchell}},\ }\href {https://doi.org/10.1038/s41467-020-19836-0} {\bibfield  {journal} {\bibinfo  {journal} {Nature Communications}\ }\textbf {\bibinfo {volume} {11}},\
  \bibinfo {pages} {6003} (\bibinfo {year} {2020})}\BibitemShut {NoStop}%
\bibitem [{\citenamefont {Li}\ \emph {et~al.}(2025)\citenamefont {Li}, \citenamefont {Gong}, \citenamefont {Zhu}, \citenamefont {Chen}, \citenamefont {Zhang}, \citenamefont {Zhang}, \citenamefont {Li}, \citenamefont {Yin}, \citenamefont {Wang}, \citenamefont {Zhao}, \citenamefont {Feng}, \citenamefont {Du},\ and\ \citenamefont {Yan}}]{yajunyan_4310}%
  \BibitemOpen
  \bibfield  {author} {\bibinfo {author} {\bibfnamefont {M.}~\bibnamefont {Li}}, \bibinfo {author} {\bibfnamefont {J.}~\bibnamefont {Gong}}, \bibinfo {author} {\bibfnamefont {Y.}~\bibnamefont {Zhu}}, \bibinfo {author} {\bibfnamefont {Z.}~\bibnamefont {Chen}}, \bibinfo {author} {\bibfnamefont {J.}~\bibnamefont {Zhang}}, \bibinfo {author} {\bibfnamefont {E.}~\bibnamefont {Zhang}}, \bibinfo {author} {\bibfnamefont {Y.}~\bibnamefont {Li}}, \bibinfo {author} {\bibfnamefont {R.}~\bibnamefont {Yin}}, \bibinfo {author} {\bibfnamefont {S.}~\bibnamefont {Wang}}, \bibinfo {author} {\bibfnamefont {J.}~\bibnamefont {Zhao}}, \bibinfo {author} {\bibfnamefont {D.-L.}\ \bibnamefont {Feng}}, \bibinfo {author} {\bibfnamefont {Z.}~\bibnamefont {Du}},\ and\ \bibinfo {author} {\bibfnamefont {Y.-J.}\ \bibnamefont {Yan}},\ }\href {https://doi.org/10.1103/2p56-xl41} {\bibfield  {journal} {\bibinfo  {journal} {Phys. Rev. B}\ }\textbf {\bibinfo {volume} {112}},\ \bibinfo {pages} {045132} (\bibinfo {year} {2025})}\BibitemShut {NoStop}%
\bibitem [{\citenamefont {Jia}\ \emph {et~al.}(2026)\citenamefont {Jia}, \citenamefont {Shen}, \citenamefont {LaBollita}, \citenamefont {Chen}, \citenamefont {Zhang}, \citenamefont {Li}, \citenamefont {Zhao}, \citenamefont {Kanatzidis}, \citenamefont {Krogstad}, \citenamefont {Zheng}, \citenamefont {Said}, \citenamefont {Alatas}, \citenamefont {Rosenkranz}, \citenamefont {Phelan}, \citenamefont {Dean}, \citenamefont {Norman}, \citenamefont {Mitchell}, \citenamefont {Botana},\ and\ \citenamefont {Cao}}]{caoyue_4310}%
  \BibitemOpen
  \bibfield  {author} {\bibinfo {author} {\bibfnamefont {X.}~\bibnamefont {Jia}}, \bibinfo {author} {\bibfnamefont {Y.}~\bibnamefont {Shen}}, \bibinfo {author} {\bibfnamefont {H.}~\bibnamefont {LaBollita}}, \bibinfo {author} {\bibfnamefont {X.}~\bibnamefont {Chen}}, \bibinfo {author} {\bibfnamefont {J.}~\bibnamefont {Zhang}}, \bibinfo {author} {\bibfnamefont {Y.}~\bibnamefont {Li}}, \bibinfo {author} {\bibfnamefont {H.}~\bibnamefont {Zhao}}, \bibinfo {author} {\bibfnamefont {M.~G.}\ \bibnamefont {Kanatzidis}}, \bibinfo {author} {\bibfnamefont {M.}~\bibnamefont {Krogstad}}, \bibinfo {author} {\bibfnamefont {H.}~\bibnamefont {Zheng}}, \bibinfo {author} {\bibfnamefont {A.~H.}\ \bibnamefont {Said}}, \bibinfo {author} {\bibfnamefont {A.}~\bibnamefont {Alatas}}, \bibinfo {author} {\bibfnamefont {S.}~\bibnamefont {Rosenkranz}}, \bibinfo {author} {\bibfnamefont {D.}~\bibnamefont {Phelan}}, \bibinfo {author} {\bibfnamefont {M.~P.~M.}\ \bibnamefont {Dean}}, \bibinfo {author} {\bibfnamefont {M.~R.}\ \bibnamefont
  {Norman}}, \bibinfo {author} {\bibfnamefont {J.~F.}\ \bibnamefont {Mitchell}}, \bibinfo {author} {\bibfnamefont {A.~S.}\ \bibnamefont {Botana}},\ and\ \bibinfo {author} {\bibfnamefont {Y.}~\bibnamefont {Cao}},\ }\href {https://doi.org/10.1103/s5j9-cbg7} {\bibfield  {journal} {\bibinfo  {journal} {Phys. Rev. X}\ }\textbf {\bibinfo {volume} {16}},\ \bibinfo {pages} {011013} (\bibinfo {year} {2026})}\BibitemShut {NoStop}%
\bibitem [{\citenamefont {Wang}\ \emph {et~al.}(2024{\natexlab{c}})\citenamefont {Wang}, \citenamefont {Jiang}, \citenamefont {Wang}, \citenamefont {Zhang},\ and\ \citenamefont {Hu}}]{Yuxin_PhysRevB.110.205122}%
  \BibitemOpen
  \bibfield  {author} {\bibinfo {author} {\bibfnamefont {Y.}~\bibnamefont {Wang}}, \bibinfo {author} {\bibfnamefont {K.}~\bibnamefont {Jiang}}, \bibinfo {author} {\bibfnamefont {Z.}~\bibnamefont {Wang}}, \bibinfo {author} {\bibfnamefont {F.-C.}\ \bibnamefont {Zhang}},\ and\ \bibinfo {author} {\bibfnamefont {J.}~\bibnamefont {Hu}},\ }\href {https://doi.org/10.1103/PhysRevB.110.205122} {\bibfield  {journal} {\bibinfo  {journal} {Phys. Rev. B}\ }\textbf {\bibinfo {volume} {110}},\ \bibinfo {pages} {205122} (\bibinfo {year} {2024}{\natexlab{c}})}\BibitemShut {NoStop}%
\bibitem [{\citenamefont {LaBollita}\ \emph {et~al.}(2024)\citenamefont {LaBollita}, \citenamefont {Pardo}, \citenamefont {Norman},\ and\ \citenamefont {Botana}}]{Botana_PhysRevMaterials.8.L111801}%
  \BibitemOpen
  \bibfield  {author} {\bibinfo {author} {\bibfnamefont {H.}~\bibnamefont {LaBollita}}, \bibinfo {author} {\bibfnamefont {V.}~\bibnamefont {Pardo}}, \bibinfo {author} {\bibfnamefont {M.~R.}\ \bibnamefont {Norman}},\ and\ \bibinfo {author} {\bibfnamefont {A.~S.}\ \bibnamefont {Botana}},\ }\href {https://doi.org/10.1103/PhysRevMaterials.8.L111801} {\bibfield  {journal} {\bibinfo  {journal} {Phys. Rev. Mater.}\ }\textbf {\bibinfo {volume} {8}},\ \bibinfo {pages} {L111801} (\bibinfo {year} {2024})}\BibitemShut {NoStop}%
\bibitem [{\citenamefont {Ni}\ \emph {et~al.}(2025)\citenamefont {Ni}, \citenamefont {Ji}, \citenamefont {He}, \citenamefont {Xie}, \citenamefont {Yao}, \citenamefont {Wang},\ and\ \citenamefont {Cao}}]{caokun_npj}%
  \BibitemOpen
  \bibfield  {author} {\bibinfo {author} {\bibfnamefont {X.-S.}\ \bibnamefont {Ni}}, \bibinfo {author} {\bibfnamefont {Y.}~\bibnamefont {Ji}}, \bibinfo {author} {\bibfnamefont {L.}~\bibnamefont {He}}, \bibinfo {author} {\bibfnamefont {T.}~\bibnamefont {Xie}}, \bibinfo {author} {\bibfnamefont {D.-X.}\ \bibnamefont {Yao}}, \bibinfo {author} {\bibfnamefont {M.}~\bibnamefont {Wang}},\ and\ \bibinfo {author} {\bibfnamefont {K.}~\bibnamefont {Cao}},\ }\href {https://doi.org/10.1038/s41535-025-00740-z} {\bibfield  {journal} {\bibinfo  {journal} {npj Quantum Materials}\ }\textbf {\bibinfo {volume} {10}},\ \bibinfo {pages} {17} (\bibinfo {year} {2025})}\BibitemShut {NoStop}%
\bibitem [{\citenamefont {Yang}\ \emph {et~al.}(2024)\citenamefont {Yang}, \citenamefont {Sun}, \citenamefont {Hu}, \citenamefont {Xie}, \citenamefont {Miao}, \citenamefont {Luo}, \citenamefont {Chen}, \citenamefont {Liang}, \citenamefont {Zhu}, \citenamefont {Qu}, \citenamefont {Chen}, \citenamefont {Huo}, \citenamefont {Huang}, \citenamefont {Zhang}, \citenamefont {Zhang}, \citenamefont {Yang}, \citenamefont {Wang}, \citenamefont {Peng}, \citenamefont {Mao}, \citenamefont {Liu}, \citenamefont {Xu}, \citenamefont {Qian}, \citenamefont {Yao}, \citenamefont {Wang}, \citenamefont {Zhao},\ and\ \citenamefont {Zhou}}]{YangJG2024}%
  \BibitemOpen
  \bibfield  {author} {\bibinfo {author} {\bibfnamefont {J.}~\bibnamefont {Yang}}, \bibinfo {author} {\bibfnamefont {H.}~\bibnamefont {Sun}}, \bibinfo {author} {\bibfnamefont {X.}~\bibnamefont {Hu}}, \bibinfo {author} {\bibfnamefont {Y.}~\bibnamefont {Xie}}, \bibinfo {author} {\bibfnamefont {T.}~\bibnamefont {Miao}}, \bibinfo {author} {\bibfnamefont {H.}~\bibnamefont {Luo}}, \bibinfo {author} {\bibfnamefont {H.}~\bibnamefont {Chen}}, \bibinfo {author} {\bibfnamefont {B.}~\bibnamefont {Liang}}, \bibinfo {author} {\bibfnamefont {W.}~\bibnamefont {Zhu}}, \bibinfo {author} {\bibfnamefont {G.}~\bibnamefont {Qu}}, \bibinfo {author} {\bibfnamefont {C.-Q.}\ \bibnamefont {Chen}}, \bibinfo {author} {\bibfnamefont {M.}~\bibnamefont {Huo}}, \bibinfo {author} {\bibfnamefont {Y.}~\bibnamefont {Huang}}, \bibinfo {author} {\bibfnamefont {S.}~\bibnamefont {Zhang}}, \bibinfo {author} {\bibfnamefont {F.}~\bibnamefont {Zhang}}, \bibinfo {author} {\bibfnamefont {F.}~\bibnamefont {Yang}}, \bibinfo {author} {\bibfnamefont
  {Z.}~\bibnamefont {Wang}}, \bibinfo {author} {\bibfnamefont {Q.}~\bibnamefont {Peng}}, \bibinfo {author} {\bibfnamefont {H.}~\bibnamefont {Mao}}, \bibinfo {author} {\bibfnamefont {G.}~\bibnamefont {Liu}}, \bibinfo {author} {\bibfnamefont {Z.}~\bibnamefont {Xu}}, \bibinfo {author} {\bibfnamefont {T.}~\bibnamefont {Qian}}, \bibinfo {author} {\bibfnamefont {D.-X.}\ \bibnamefont {Yao}}, \bibinfo {author} {\bibfnamefont {M.}~\bibnamefont {Wang}}, \bibinfo {author} {\bibfnamefont {L.}~\bibnamefont {Zhao}},\ and\ \bibinfo {author} {\bibfnamefont {X.~J.}\ \bibnamefont {Zhou}},\ }\href {https://doi.org/10.1038/s41467-024-48701-7} {\bibfield  {journal} {\bibinfo  {journal} {Nature Communications}\ }\textbf {\bibinfo {volume} {15}},\ \bibinfo {pages} {4373} (\bibinfo {year} {2024})}\BibitemShut {NoStop}%
\bibitem [{\citenamefont {Li}\ \emph {et~al.}(2017)\citenamefont {Li}, \citenamefont {Zhou}, \citenamefont {Nummy}, \citenamefont {Zhang}, \citenamefont {Pardo}, \citenamefont {Pickett}, \citenamefont {Mitchell},\ and\ \citenamefont {Dessau}}]{lihaoxiang}%
  \BibitemOpen
  \bibfield  {author} {\bibinfo {author} {\bibfnamefont {H.}~\bibnamefont {Li}}, \bibinfo {author} {\bibfnamefont {X.}~\bibnamefont {Zhou}}, \bibinfo {author} {\bibfnamefont {T.}~\bibnamefont {Nummy}}, \bibinfo {author} {\bibfnamefont {J.}~\bibnamefont {Zhang}}, \bibinfo {author} {\bibfnamefont {V.}~\bibnamefont {Pardo}}, \bibinfo {author} {\bibfnamefont {W.~E.}\ \bibnamefont {Pickett}}, \bibinfo {author} {\bibfnamefont {J.~F.}\ \bibnamefont {Mitchell}},\ and\ \bibinfo {author} {\bibfnamefont {D.~S.}\ \bibnamefont {Dessau}},\ }\href {https://doi.org/10.1038/s41467-017-00777-0} {\bibfield  {journal} {\bibinfo  {journal} {Nature Communications}\ }\textbf {\bibinfo {volume} {8}},\ \bibinfo {pages} {704} (\bibinfo {year} {2017})}\BibitemShut {NoStop}%
\bibitem [{\citenamefont {Du}\ \emph {et~al.}(2025)\citenamefont {Du}, \citenamefont {Wang}, \citenamefont {Li}, \citenamefont {Cao}, \citenamefont {Zhang}, \citenamefont {Pei}, \citenamefont {Yang}, \citenamefont {Zhao}, \citenamefont {Zhai}, \citenamefont {Liu}, \citenamefont {Li}, \citenamefont {Zhao}, \citenamefont {Liu}, \citenamefont {Shen}, \citenamefont {Li}, \citenamefont {He}, \citenamefont {Chen}, \citenamefont {Qi}, \citenamefont {Guo},\ and\ \citenamefont {Yang}}]{yanglexian}%
  \BibitemOpen
  \bibfield  {author} {\bibinfo {author} {\bibfnamefont {X.}~\bibnamefont {Du}}, \bibinfo {author} {\bibfnamefont {Y.~L.}\ \bibnamefont {Wang}}, \bibinfo {author} {\bibfnamefont {Y.~D.}\ \bibnamefont {Li}}, \bibinfo {author} {\bibfnamefont {Y.~T.}\ \bibnamefont {Cao}}, \bibinfo {author} {\bibfnamefont {M.~X.}\ \bibnamefont {Zhang}}, \bibinfo {author} {\bibfnamefont {C.~Y.}\ \bibnamefont {Pei}}, \bibinfo {author} {\bibfnamefont {J.~M.}\ \bibnamefont {Yang}}, \bibinfo {author} {\bibfnamefont {W.~X.}\ \bibnamefont {Zhao}}, \bibinfo {author} {\bibfnamefont {K.~Y.}\ \bibnamefont {Zhai}}, \bibinfo {author} {\bibfnamefont {Z.~K.}\ \bibnamefont {Liu}}, \bibinfo {author} {\bibfnamefont {Z.~W.}\ \bibnamefont {Li}}, \bibinfo {author} {\bibfnamefont {J.~K.}\ \bibnamefont {Zhao}}, \bibinfo {author} {\bibfnamefont {Z.~T.}\ \bibnamefont {Liu}}, \bibinfo {author} {\bibfnamefont {D.~W.}\ \bibnamefont {Shen}}, \bibinfo {author} {\bibfnamefont {Z.}~\bibnamefont {Li}}, \bibinfo {author} {\bibfnamefont {Y.}~\bibnamefont {He}},
  \bibinfo {author} {\bibfnamefont {Y.~L.}\ \bibnamefont {Chen}}, \bibinfo {author} {\bibfnamefont {Y.~P.}\ \bibnamefont {Qi}}, \bibinfo {author} {\bibfnamefont {H.~J.}\ \bibnamefont {Guo}},\ and\ \bibinfo {author} {\bibfnamefont {L.~X.}\ \bibnamefont {Yang}},\ }\href {https://doi.org/10.1103/smkf-k7wq} {\bibfield  {journal} {\bibinfo  {journal} {Phys. Rev. Lett.}\ }\textbf {\bibinfo {volume} {135}},\ \bibinfo {pages} {146506} (\bibinfo {year} {2025})}\BibitemShut {NoStop}%
\bibitem [{\citenamefont {Yang}\ \emph {et~al.}(2026)\citenamefont {Yang}, \citenamefont {Zhan}, \citenamefont {Miao}, \citenamefont {Huo}, \citenamefont {Xu}, \citenamefont {Li}, \citenamefont {Xie}, \citenamefont {Liang}, \citenamefont {Cai}, \citenamefont {Chen}, \citenamefont {Zhu}, \citenamefont {Xu}, \citenamefont {Zhang}, \citenamefont {Zhang}, \citenamefont {Yang}, \citenamefont {Wang}, \citenamefont {Peng}, \citenamefont {Mao}, \citenamefont {Li}, \citenamefont {Zhu}, \citenamefont {Liu}, \citenamefont {Xu}, \citenamefont {Hu}, \citenamefont {Wu}, \citenamefont {Wang}, \citenamefont {Zhao},\ and\ \citenamefont {Zhou}}]{yang2026electronicorigindensitywave}%
  \BibitemOpen
  \bibfield  {author} {\bibinfo {author} {\bibfnamefont {J.}~\bibnamefont {Yang}}, \bibinfo {author} {\bibfnamefont {J.}~\bibnamefont {Zhan}}, \bibinfo {author} {\bibfnamefont {T.}~\bibnamefont {Miao}}, \bibinfo {author} {\bibfnamefont {M.}~\bibnamefont {Huo}}, \bibinfo {author} {\bibfnamefont {Q.}~\bibnamefont {Xu}}, \bibinfo {author} {\bibfnamefont {Y.}~\bibnamefont {Li}}, \bibinfo {author} {\bibfnamefont {Y.}~\bibnamefont {Xie}}, \bibinfo {author} {\bibfnamefont {B.}~\bibnamefont {Liang}}, \bibinfo {author} {\bibfnamefont {N.}~\bibnamefont {Cai}}, \bibinfo {author} {\bibfnamefont {H.}~\bibnamefont {Chen}}, \bibinfo {author} {\bibfnamefont {W.}~\bibnamefont {Zhu}}, \bibinfo {author} {\bibfnamefont {M.}~\bibnamefont {Xu}}, \bibinfo {author} {\bibfnamefont {S.}~\bibnamefont {Zhang}}, \bibinfo {author} {\bibfnamefont {F.}~\bibnamefont {Zhang}}, \bibinfo {author} {\bibfnamefont {F.}~\bibnamefont {Yang}}, \bibinfo {author} {\bibfnamefont {Z.}~\bibnamefont {Wang}}, \bibinfo {author} {\bibfnamefont
  {Q.}~\bibnamefont {Peng}}, \bibinfo {author} {\bibfnamefont {H.}~\bibnamefont {Mao}}, \bibinfo {author} {\bibfnamefont {X.}~\bibnamefont {Li}}, \bibinfo {author} {\bibfnamefont {Z.}~\bibnamefont {Zhu}}, \bibinfo {author} {\bibfnamefont {G.}~\bibnamefont {Liu}}, \bibinfo {author} {\bibfnamefont {Z.}~\bibnamefont {Xu}}, \bibinfo {author} {\bibfnamefont {J.}~\bibnamefont {Hu}}, \bibinfo {author} {\bibfnamefont {X.}~\bibnamefont {Wu}}, \bibinfo {author} {\bibfnamefont {M.}~\bibnamefont {Wang}}, \bibinfo {author} {\bibfnamefont {L.}~\bibnamefont {Zhao}},\ and\ \bibinfo {author} {\bibfnamefont {X.~J.}\ \bibnamefont {Zhou}},\ }\href {https://arxiv.org/abs/2601.22608} {\bibinfo {title} {Electronic origin of density wave orders in a trilayer nickelate}} (\bibinfo {year} {2026}),\ \Eprint {https://arxiv.org/abs/2601.22608} {arXiv:2601.22608 [cond-mat.supr-con]} \BibitemShut {NoStop}%
\bibitem [{\citenamefont {Jiang}\ \emph {et~al.}(2026)\citenamefont {Jiang}, \citenamefont {Zhang}, \citenamefont {Wang}, \citenamefont {Liu}, \citenamefont {Liu}, \citenamefont {Zhang}, \citenamefont {Zhang}, \citenamefont {Jing}, \citenamefont {Huang}, \citenamefont {Jiang}, \citenamefont {Ye}, \citenamefont {Jiang}, \citenamefont {Zhao}, \citenamefont {Shen},\ and\ \citenamefont {Feng}}]{daweishen_4310}%
  \BibitemOpen
  \bibfield  {author} {\bibinfo {author} {\bibfnamefont {Z.}~\bibnamefont {Jiang}}, \bibinfo {author} {\bibfnamefont {E.}~\bibnamefont {Zhang}}, \bibinfo {author} {\bibfnamefont {Y.}~\bibnamefont {Wang}}, \bibinfo {author} {\bibfnamefont {Z.}~\bibnamefont {Liu}}, \bibinfo {author} {\bibfnamefont {J.}~\bibnamefont {Liu}}, \bibinfo {author} {\bibfnamefont {R.}~\bibnamefont {Zhang}}, \bibinfo {author} {\bibfnamefont {X.}~\bibnamefont {Zhang}}, \bibinfo {author} {\bibfnamefont {W.}~\bibnamefont {Jing}}, \bibinfo {author} {\bibfnamefont {Y.}~\bibnamefont {Huang}}, \bibinfo {author} {\bibfnamefont {Q.}~\bibnamefont {Jiang}}, \bibinfo {author} {\bibfnamefont {M.}~\bibnamefont {Ye}}, \bibinfo {author} {\bibfnamefont {K.}~\bibnamefont {Jiang}}, \bibinfo {author} {\bibfnamefont {J.}~\bibnamefont {Zhao}}, \bibinfo {author} {\bibfnamefont {D.}~\bibnamefont {Shen}},\ and\ \bibinfo {author} {\bibfnamefont {D.}~\bibnamefont {Feng}},\ }\href {https://arxiv.org/abs/2602.02127} {\bibinfo {title} {Direct observation of
  unidirectional density wave and band splitting in a single-domain trilayer nickelate {Pr$_4$Ni$_3$O$_{10}$}}} (\bibinfo {year} {2026}),\ \Eprint {https://arxiv.org/abs/2602.02127} {arXiv:2602.02127 [cond-mat.supr-con]} \BibitemShut {NoStop}%
\bibitem [{\citenamefont {Brydon}\ and\ \citenamefont {Timm}(2009)}]{iron_PhysRevB.80.174401}%
  \BibitemOpen
  \bibfield  {author} {\bibinfo {author} {\bibfnamefont {P.~M.~R.}\ \bibnamefont {Brydon}}\ and\ \bibinfo {author} {\bibfnamefont {C.}~\bibnamefont {Timm}},\ }\href {https://doi.org/10.1103/PhysRevB.80.174401} {\bibfield  {journal} {\bibinfo  {journal} {Phys. Rev. B}\ }\textbf {\bibinfo {volume} {80}},\ \bibinfo {pages} {174401} (\bibinfo {year} {2009})}\BibitemShut {NoStop}%
\bibitem [{\citenamefont {Knolle}\ \emph {et~al.}(2010)\citenamefont {Knolle}, \citenamefont {Eremin}, \citenamefont {Chubukov},\ and\ \citenamefont {Moessner}}]{iron_PhysRevB.81.140506}%
  \BibitemOpen
  \bibfield  {author} {\bibinfo {author} {\bibfnamefont {J.}~\bibnamefont {Knolle}}, \bibinfo {author} {\bibfnamefont {I.}~\bibnamefont {Eremin}}, \bibinfo {author} {\bibfnamefont {A.~V.}\ \bibnamefont {Chubukov}},\ and\ \bibinfo {author} {\bibfnamefont {R.}~\bibnamefont {Moessner}},\ }\href {https://doi.org/10.1103/PhysRevB.81.140506} {\bibfield  {journal} {\bibinfo  {journal} {Phys. Rev. B}\ }\textbf {\bibinfo {volume} {81}},\ \bibinfo {pages} {140506} (\bibinfo {year} {2010})}\BibitemShut {NoStop}%
\bibitem [{\citenamefont {Kaneshita}\ and\ \citenamefont {Tohyama}(2010)}]{iron_PhysRevB.82.094441}%
  \BibitemOpen
  \bibfield  {author} {\bibinfo {author} {\bibfnamefont {E.}~\bibnamefont {Kaneshita}}\ and\ \bibinfo {author} {\bibfnamefont {T.}~\bibnamefont {Tohyama}},\ }\href {https://doi.org/10.1103/PhysRevB.82.094441} {\bibfield  {journal} {\bibinfo  {journal} {Phys. Rev. B}\ }\textbf {\bibinfo {volume} {82}},\ \bibinfo {pages} {094441} (\bibinfo {year} {2010})}\BibitemShut {NoStop}%
\bibitem [{\citenamefont {Knolle}\ \emph {et~al.}(2011)\citenamefont {Knolle}, \citenamefont {Eremin},\ and\ \citenamefont {Moessner}}]{iron_PhysRevB.83.224503}%
  \BibitemOpen
  \bibfield  {author} {\bibinfo {author} {\bibfnamefont {J.}~\bibnamefont {Knolle}}, \bibinfo {author} {\bibfnamefont {I.}~\bibnamefont {Eremin}},\ and\ \bibinfo {author} {\bibfnamefont {R.}~\bibnamefont {Moessner}},\ }\href {https://doi.org/10.1103/PhysRevB.83.224503} {\bibfield  {journal} {\bibinfo  {journal} {Phys. Rev. B}\ }\textbf {\bibinfo {volume} {83}},\ \bibinfo {pages} {224503} (\bibinfo {year} {2011})}\BibitemShut {NoStop}%
\bibitem [{\citenamefont {Norman}(2025)}]{norman-landau}%
  \BibitemOpen
  \bibfield  {author} {\bibinfo {author} {\bibfnamefont {M.~R.}\ \bibnamefont {Norman}},\ }\href {https://doi.org/10.1103/43qs-x65n} {\bibfield  {journal} {\bibinfo  {journal} {Phys. Rev. B}\ }\textbf {\bibinfo {volume} {112}},\ \bibinfo {pages} {075149} (\bibinfo {year} {2025})}\BibitemShut {NoStop}%
\bibitem [{Sup()}]{SuppMat}%
  \BibitemOpen
  \href@noop {} {}\bibinfo {note} {See Supplemental Material for: (i) tight-binding models, bandwidth rescalings, and mirror-basis/folded-momentum conventions; (ii) the residual-interaction model and Hartree--Fock/RPA formulation; (iii) orbital/layer projections of the ordered spin and charge textures, including the induced $2\mathbf Q$ charge modulation; and (iv) a local-spin comparison for large interlayer exchange.}\BibitemShut {Stop}%
\bibitem [{\citenamefont {Kaneshita}\ \emph {et~al.}(2001)\citenamefont {Kaneshita}, \citenamefont {Ichioka},\ and\ \citenamefont {Machida}}]{doi:10.1143/JPSJ.70.866}%
  \BibitemOpen
  \bibfield  {author} {\bibinfo {author} {\bibfnamefont {E.}~\bibnamefont {Kaneshita}}, \bibinfo {author} {\bibfnamefont {M.}~\bibnamefont {Ichioka}},\ and\ \bibinfo {author} {\bibfnamefont {K.}~\bibnamefont {Machida}},\ }\href {https://doi.org/10.1143/JPSJ.70.866} {\bibfield  {journal} {\bibinfo  {journal} {Journal of the Physical Society of Japan}\ }\textbf {\bibinfo {volume} {70}},\ \bibinfo {pages} {866} (\bibinfo {year} {2001})}\BibitemShut {NoStop}%
\end{thebibliography}%

\enabletocentries

\clearpage
\onecolumngrid


\begin{center}
{\large\bf Supplemental Material for}\\[0.5em]
{\large\bf ``Itinerant Nature of Spin-Density-Wave Order in Ruddlesden--Popper Nickelates''}
\end{center}

\vspace{1em}

\setcounter{tocdepth}{2}
\setcounter{secnumdepth}{2}

\setcounter{section}{0}
\setcounter{subsection}{0}
\setcounter{equation}{0}
\setcounter{figure}{0}
\setcounter{table}{0}

\renewcommand{\thesection}{\Roman{section}}
\renewcommand{\thesubsection}{\Alph{subsection}}

\renewcommand{\theequation}{S\arabic{equation}}
\renewcommand{\thefigure}{S\arabic{figure}}
\renewcommand{\thetable}{S\arabic{table}}

\renewcommand{\theHequation}{S\arabic{equation}}
\renewcommand{\theHfigure}{S\arabic{figure}}
\renewcommand{\theHtable}{S\arabic{table}}
\renewcommand{\theHsection}{S\arabic{section}}
\renewcommand{\theHsubsection}{S\arabic{section}.\arabic{subsection}}

\tableofcontents

\section{Tight-binding models and mirror-band basis}

Throughout this Supplemental Material, LP327 and LP4310 denote the
low-pressure \LNOtwoseven\ and \LNOfourten\ settings considered in the main
text, respectively.

\subsection{Bilayer \LNOtwoseven}

The bilayer calculation starts from the low-pressure eight-orbital
tight-binding model of Ref.~\cite{Yuxin_PhysRevB.110.205122}.
\begin{equation}
C^{327}_{\bk\sigma}
=
\left(
c_{1,1,x},
c_{1,1,z},
c_{1,2,x},
c_{1,2,z},
c_{2,1,x},
c_{2,1,z},
c_{2,2,x},
c_{2,2,z}
\right)_{\bk\sigma}^{T}.
\label{eq:lp327_basis}
\end{equation}
The first index labels the layer, the second labels the two in-plane Ni sites
in the low-pressure unit cell, and $x,z$ denote the
$d_{x^2-y^2}$ and $d_{z^2}$ orbitals.  In this basis
\begin{equation}
H^{327}_0(\bk)
=
\begin{pmatrix}
H_t(\bk) & H_{\perp}(\bk)\\
H_{\perp}^{\dagger}(\bk) & H_t(\bk)
\end{pmatrix},
\label{eq:lp327_h0_block}
\end{equation}
where the intralayer and interlayer blocks are
\begin{align}
H_t(\bk)
&=
\begin{pmatrix}
H_{11} & 0 & H_{13} & H_{14}\\
0 & H_{22} & H_{14} & H_{24}\\
H_{13}^{*} & H_{14}^{*} & H_{11} & 0\\
H_{14}^{*} & H_{24}^{*} & 0 & H_{22}
\end{pmatrix},
\nonumber\\
H_{\perp}(\bk)
&=
\begin{pmatrix}
t_{vx} & 0 & 0 & H_{18}\\
0 & t_{vz} & H_{18} & 0\\
0 & H_{18}^{*} & t_{vx} & 0\\
H_{18}^{*} & 0 & 0 & t_{vz}
\end{pmatrix}.
\end{align}
The momentum-dependent matrix elements are
\begin{align}
H_{11}(\bk)
&=
e_x+2\left(t_{2x}'\cos k_x+t_{2x}\cos k_y\right)
+4t_{5x}\cos k_x\cos k_y,
\nonumber\\
H_{22}(\bk)
&=
e_z+2\left(t_{2z}'\cos k_x+t_{2z}\cos k_y\right)
+4t_{5z}\cos k_x\cos k_y,
\nonumber\\
H_{13}(\bk)
&=
2\cos\frac{k_x}{2}
\left(t_{1x}\ee^{\ii k_y/2}+t_{1x}'\ee^{-\ii k_y/2}\right),
\nonumber\\
H_{24}(\bk)
&=
2\cos\frac{k_x}{2}
\left(t_{1z}\ee^{\ii k_y/2}+t_{1z}'\ee^{-\ii k_y/2}\right),
\nonumber\\
H_{14}(\bk)
&=
2\ii\sin\frac{k_x}{2}
\left(t_{3xz}\ee^{\ii k_y/2}-t_{3xz}'\ee^{-\ii k_y/2}\right),
\nonumber\\
H_{18}(\bk)
&=
2\ii\sin\frac{k_x}{2}
\left(t_{4xz}\ee^{\ii k_y/2}-t_{4xz}'\ee^{-\ii k_y/2}\right).
\label{eq:lp327_h_elements}
\end{align}
The hopping parameters are listed in Table~\ref{tab:lp327_tb}.  All energies
are in eV.

\begin{table}[htbp]
\centering
\begin{tabular}{@{}lr@{\hspace{3em}}lr@{}}
parameter & value & parameter & value\\
\hline
$t_{1x}$ & $-0.5192$ & $t_{1x}'$ & $-0.5253$\\
$t_{1z}$ & $-0.1232$ & $t_{1z}'$ & $-0.1038$\\
$t_{2x}$ & $0.0693$ & $t_{2x}'$ & $0.0804$\\
$t_{2z}$ & $-0.0185$ & $t_{2z}'$ & $-0.0179$\\
$t_{3xz}$ & $0.2712$ & $t_{3xz}'$ & $0.2577$\\
$t_{4xz}$ & $-0.0332$ & $t_{4xz}'$ & $-0.0255$\\
$t_{5x}$ & $-0.0509$ & $t_{5z}$ & $-0.0131$\\
$t_{vx}$ & $0.0173$ & $t_{vz}$ & $-0.8351$\\
$e_x$ & $1.1957$ & $e_z$ & $0.0393$
\end{tabular}
\caption{Tight-binding parameters for the bilayer \LNOtwoseven\ model, adapted
from Ref.~\cite{Yuxin_PhysRevB.110.205122}.}
\label{tab:lp327_tb}
\end{table}

Mirror reflection exchanges the two layers and leaves the in-plane momentum
unchanged:
\begin{equation}
M_z^{327}
=
\begin{pmatrix}
0_4 & I_4\\
I_4 & 0_4
\end{pmatrix},
\qquad
U_M^{327}
=
\frac{1}{\sqrt 2}
\begin{pmatrix}
I_4 & I_4\\
I_4 & -I_4
\end{pmatrix}.
\label{eq:lp327_mirror}
\end{equation}
The mirror-band operators are
\begin{equation}
\widetilde C^{327}_{\bk\sigma}=(U_M^{327})^\dagger C^{327}_{\bk\sigma},
\qquad
(U_M^{327})^\dagger H^{327}_0(\bk)U_M^{327}
=
\begin{pmatrix}
H_+(\bk)&0\\
0&H_-(\bk)
\end{pmatrix}.
\label{eq:lp327_mirror_blocks}
\end{equation}
Within each mirror sector the bands are ordered by increasing eigenvalue.
In the notation of the main text, the LP327 Fermi-level manifold contains
three mirror-resolved bands,
\begin{equation}
\alpha,\qquad \beta,\qquad \beta^\prime .
\label{eq:lp327_retained}
\end{equation}
These three bands define the explicit low-energy one-particle subspace.  For
the mirror-odd SDW calculation discussed in the main text, the residual
interaction is restricted to the $\alpha$--$\beta$ channel: $g_{\alpha\beta}$
and $g_{\alpha\beta,1}$ are kept, whereas the $\alpha$--$\beta^\prime$ and
$\beta$--$\beta^\prime$ residual vertices are set to zero.  Bands outside this
Fermi-level manifold carry no residual interaction vertices and enter only as
noninteracting background bands that fix the chemical potential.

\subsection{Trilayer \LNOfourten}

For the trilayer material we use the high-pressure three-layer cell as a
mirror-symmetric representation of the low-pressure electronic structure.  The
corresponding tight-binding parametrization is adapted from
Ref.~\cite{yang2026electronicorigindensitywave}.
The one-spin basis is
\begin{equation}
C^{4310}_{\bk\sigma}
=
\left(t_x,t_z,m_x,m_z,b_x,b_z\right)_{\bk\sigma}^{T},
\label{eq:lp4310_basis}
\end{equation}
where $t,m,b$ denote top, middle, and bottom NiO$_2$ layers.  The Hamiltonian
is written in $2\times2$ layer blocks,
\begin{equation}
H^{4310}_0(\bk)
=
\begin{pmatrix}
H_t & H_{tm} & H_{tb}\\
H_{tm}^{\dagger} & H_m & H_{mb}\\
H_{tb}^{\dagger} & H_{mb}^{\dagger} & H_b
\end{pmatrix},
\qquad
H_b=H_t,\quad H_{mb}=H_{tm}^{T}.
\label{eq:lp4310_h0_block}
\end{equation}
Let
\begin{equation}
c_+(\bk)=\cos k_x+\cos k_y,\qquad
c_{xy}(\bk)=\cos k_x\cos k_y,\qquad
d_-(\bk)=\cos k_x-\cos k_y .
\label{eq:lp4310_form_factors}
\end{equation}
The layer blocks are
\begin{align}
H_t&=
\begin{pmatrix}
e_{x,t}+2t_{Nx,t}c_+ +4t_{NNx,t}c_{xy}
&
2t_{Nxz,t}d_-
\\
2t_{Nxz,t}d_-
&
e_{z,t}+2t_{Nz,t}c_+ +4t_{NNz,t}c_{xy}
\end{pmatrix},
\nonumber\\
H_m&=
\begin{pmatrix}
e_{x,m}+2t_{Nx,m}c_+ +4t_{NNx,m}c_{xy}
&
2t_{Nxz,m}d_-
\\
2t_{Nxz,m}d_-
&
e_{z,m}+2t_{Nz,m}c_+ +4t_{NNz,m}c_{xy}
\end{pmatrix},
\nonumber\\
H_{tm}&=
\begin{pmatrix}
t_{\perp x}+2s_{Nx}c_+ +4s_{NNx}c_{xy}
&
2s_{Nxz,1}d_-
\\
2s_{Nxz,2}d_-
&
t_{\perp z}+2s_{Nz}c_+ +4s_{NNz}c_{xy}
\end{pmatrix},
\nonumber\\
H_{tb}&=
\begin{pmatrix}
t_{\perp2,x} & 0\\
0 & t_{\perp2,z}
\end{pmatrix}.
\label{eq:lp4310_h_blocks}
\end{align}
The numerical parameters are given in Table~\ref{tab:lp4310_tb}.

\begin{table}[htbp]
\centering
\begin{tabular}{@{}lr@{\hspace{3em}}lr@{}}
parameter & value & parameter & value\\
\hline
$e_{x,m}$ & $1.0518$ & $e_{x,t}$ & $0.8666$\\
$e_{z,m}$ & $0.6000$ & $e_{z,t}$ & $0.8000$\\
$s_{NNx}$ & $0.0184$ & $s_{NNz}$ & $-0.0245$\\
$s_{Nx}$ & $0.0339$ & $s_{Nz}$ & $0.0262$\\
$s_{Nxz,1}$ & $0.0294$ & $s_{Nxz,2}$ & $0.0597$\\
$t_{NNx,m}$ & $-0.1000$ & $t_{NNx,t}$ & $0.0830$\\
$t_{NNz,m}$ & $0.0369$ & $t_{NNz,t}$ & $-0.0111$\\
$t_{Nx,m}$ & $-0.5165$ & $t_{Nx,t}$ & $-0.5248$\\
$t_{Nxz,m}$ & $-0.3642$ & $t_{Nxz,t}$ & $-0.2752$\\
$t_{Nz,m}$ & $-0.1798$ & $t_{Nz,t}$ & $-0.1409$\\
$t_{\perp2,x}$ & $-0.0550$ & $t_{\perp2,z}$ & $0.0192$\\
$t_{\perp x}$ & $-0.0275$ & $t_{\perp z}$ & $-0.7503$
\end{tabular}
\caption{Tight-binding parameters for the trilayer \LNOfourten\ model, adapted
from Ref.~\cite{yang2026electronicorigindensitywave}.}
\label{tab:lp4310_tb}
\end{table}

The mirror operation exchanges the two outer layers.  A convenient
mirror-adapted basis is
\begin{equation}
\left(
\frac{t_x+b_x}{\sqrt2},
\frac{t_z+b_z}{\sqrt2},
m_x,
m_z;
\frac{t_x-b_x}{\sqrt2},
\frac{t_z-b_z}{\sqrt2}
\right),
\label{eq:lp4310_mirror_basis}
\end{equation}
where the first four states are mirror even and the last two are mirror odd.
The Fermi-level nesting channel retained in the interacting calculation
connects the mirror-even $\alpha$ band and the mirror-odd $\beta^\prime$
band, as in the main text.

\section{Mirror-resolved susceptibility and nesting}

The nesting analysis in the main text uses the normal-state band-resolved
susceptibility $\chi_{\rm N}^{ab}(\bq)$.  For two mirror bands $a$ and $b$,
\begin{equation}
\chi_{\rm N}^{ab}(\bq)
=
-\frac{1}{N_{\rm BZ}}\sum_{\bk\in\mathrm{BZ}}
\frac{
f[\xi_a(\bk+\bq)]-f[\xi_b(\bk)]
}{
\xi_a(\bk+\bq)-\xi_b(\bk)+\ii0^+
}.
\label{eq:band_resolved_chi}
\end{equation}
Here $N_{\rm BZ}$ denotes the number of momenta in the full Brillouin-zone
mesh, $\xi_\lambda(\bk)=\epsilon_\lambda(\bk)-\mu$, $\lambda$ labels a
mirror-resolved band of $H_0$, and $f$ is the Fermi-Dirac distribution function.  Since the one-particle Hamiltonians in
Eqs.~\eqref{eq:lp327_mirror_blocks} and \eqref{eq:lp4310_mirror_basis}
commute with mirror reflection, interband bubbles may be classified by the
product of mirror eigenvalues.  Opposite-parity channels generate a mirror-odd
interband SDW.

For LP327 the main-text susceptibility is
\begin{equation}
\chi_{\rm N}^{\alpha\beta}(\bq),
\label{eq:lp327_chi_ab}
\end{equation}
whose maximum lies near $\bQ_1\simeq(0,\pm1.16\pi)$.  The ordered-state
calculation uses the nearby commensurate vector $\bQ=(0,\pi)$, which is
self-conjugate in the sense that $-\bQ\equiv\bQ$ modulo a reciprocal lattice
vector, or equivalently $2\bQ\equiv0$. For LP4310 the corresponding channel is
\begin{equation}
\chi_{\rm N}^{\alpha\beta^\prime}(\bq),
\label{eq:lp4310_chi_abp}
\end{equation}
with a maximum near $(0.62\pi,0.62\pi)$; the commensurate Hartree-Fock/RPA
calculation uses $\bQ=(2\pi/3,2\pi/3)$.

\section{Effective interaction in the retained bands}

We use the same band-basis interaction notation as the main text.  For two
retained bands $a$ and $b$, the five local four-fermion vertices are
\begin{equation}
H_I^{ab}
=
H_{aa}+H_{bb}+H_{ab}+H_{ab,1}+H_{ab,2},
\label{eq:five_terms_main_notation}
\end{equation}
with
\begin{align}
H_{aa}
&=
\frac{g_{aa}}{N_{\rm BZ}}
\sum_{\bk,\bk',\bq}
a^\dagger_{\bk+\bq,\uparrow}a_{\bk,\uparrow}
a^\dagger_{\bk'-\bq,\downarrow}a_{\bk',\downarrow},
\nonumber\\
H_{bb}
&=
\frac{g_{bb}}{N_{\rm BZ}}
\sum_{\bk,\bk',\bq}
b^\dagger_{\bk+\bq,\uparrow}b_{\bk,\uparrow}
b^\dagger_{\bk'-\bq,\downarrow}b_{\bk',\downarrow},
\nonumber\\
H_{ab}
&=
\frac{g_{ab}}{N_{\rm BZ}}
\sum_{\bk,\bk',\bq}\sum_{\sigma,\sigma'}
a^\dagger_{\bk+\bq,\sigma}a_{\bk,\sigma}
b^\dagger_{\bk'-\bq,\sigma'}b_{\bk',\sigma'},
\nonumber\\
H_{ab,1}
&=
\frac{g_{ab,1}}{N_{\rm BZ}}
\sum_{\bk,\bk',\bq}
\left(
a^\dagger_{\bk+\bq,\uparrow}a^\dagger_{\bk'-\bq,\downarrow}
b_{\bk',\downarrow}b_{\bk,\uparrow}
+{\rm H.c.}
\right),
\nonumber\\
H_{ab,2}
&=
\frac{g_{ab,2}}{N_{\rm BZ}}
\sum_{\bk,\bk',\bq}\sum_{\sigma,\sigma'}
a^\dagger_{\bk+\bq,\sigma}b^\dagger_{\bk'-\bq,\sigma'}
a_{\bk',\sigma'}b_{\bk,\sigma}.
\label{eq:five_interactions_main_notation}
\end{align}
The momenta in Eq.~\eqref{eq:five_interactions_main_notation} are full
Brillouin-zone momenta.  After folding into an ordered commensurate problem,
we write the corresponding normalization as $N_{\rm full}=N N_{\RBZ}$, with
$N=2$ for LP327 and $N=3$ for LP4310.
This is the interband SDW interaction structure used in the main text and in
standard itinerant SDW treatments~\cite{iron_PhysRevB.80.174401,iron_PhysRevB.83.224503}.

For LP327 the active SDW pair is
\begin{equation}
a=\alpha,\qquad b=\beta .
\label{eq:lp327_active_pair}
\end{equation}
Only $g_{\alpha\beta}$ and $g_{\alpha\beta,1}$ are nonzero in the main-text
calculation.  Residual vertices involving $\beta^\prime$ are set to zero.
The $\beta^\prime$ band is nevertheless retained in the low-energy
one-particle spectrum and in the filling constraint, as stated after
Eq.~\eqref{eq:lp327_retained}.

For LP4310 the active pair is
\begin{equation}
a=\alpha,\qquad b=\beta^\prime ,
\label{eq:lp4310_active_pair}
\end{equation}
and the five couplings are those listed in the main text for the
$\alpha$--$\beta^\prime$ sector.

The transverse RPA vertex is defined in the same convention as the bare
bubble.  With pair index
\begin{equation}
I=(l_2-1)N_b+l_1,
\qquad
l_1,l_2\in\{a,b\},
\qquad
N_b=2,
\label{eq:pair_index_main}
\end{equation}
the pair order is
\begin{equation}
(aa,ba,ab,bb).
\label{eq:pair_order_main}
\end{equation}
In this order the local transverse vertex is
\begin{equation}
U^{+-}_{\rm pair}
=
\begin{pmatrix}
g_{aa} & 0 & 0 & g_{ab,2}\\
0 & g_{ab} & g_{ab,1} & 0\\
0 & g_{ab,1} & g_{ab} & 0\\
g_{ab,2} & 0 & 0 & g_{bb}
\end{pmatrix}_{(aa,ba,ab,bb)}.
\label{eq:upair_main}
\end{equation}
If additional spectator bands are retained in $H_0$, all residual-vertex
matrix elements involving those bands are set to zero.

\section{Hartree-Fock SDW formulation}

\subsection{Self-conjugate LP327 case}

For LP327 we use the commensurate ansatz $\bQ=(0,\pi)$, for which
$2\bQ$ is a reciprocal lattice vector.  For the numerical Hartree-Fock/RPA
calculation, the one-particle energies entering Eq.~\eqref{eq:lp327_hf_matrix}
are multiplied by a global bandwidth rescaling factor of $0.15$, motivated by
ARPES measurements~\cite{YangJG2024}.  The retained low-energy folded spinor is
\begin{equation}
\Psi^{327}_{\bk\sigma}
=
\left(
\alpha_{\bk\sigma},
\beta_{\bk\sigma},
\beta^\prime_{\bk\sigma},
\alpha_{\bk+\bQ,\sigma},
\beta_{\bk+\bQ,\sigma},
\beta^\prime_{\bk+\bQ,\sigma}
\right)^T,
\qquad \bk\in\RBZ .
\label{eq:lp327_folded_spinor}
\end{equation}
The main-text SDW channel is the self-conjugate $\alpha$--$\beta$ channel.
With all residual vertices involving $\beta^\prime$ set to zero, the
mean-field Hamiltonian is
\begin{equation}
\mathcal H^{327}_{\sigma}(\bk)
=
\begin{pmatrix}
\xi^\alpha_{\bk}-\delta\mu & 0 & 0 & 0 & \sigma\Delta_{\alpha\beta} & 0\\
0 & \xi^\beta_{\bk}-\delta\mu & 0 & \sigma\Delta_{\alpha\beta} & 0 & 0\\
0 & 0 & \xi^{\beta^\prime}_{\bk}-\delta\mu & 0 & 0 & 0\\
0 & \sigma\Delta_{\alpha\beta} & 0 & \xi^\alpha_{\bk+\bQ}-\delta\mu & 0 & 0\\
\sigma\Delta_{\alpha\beta} & 0 & 0 & 0 & \xi^\beta_{\bk+\bQ}-\delta\mu & 0\\
0 & 0 & 0 & 0 & 0 & \xi^{\beta^\prime}_{\bk+\bQ}-\delta\mu
\end{pmatrix}.
\label{eq:lp327_hf_matrix}
\end{equation}
With $N_{\rm full}^{327}=2N_{\RBZ}$, the normalized self-conjugate SDW
bilinear is
\begin{equation}
M_{\alpha\beta}^{327}
=
\frac{1}{N_{\rm full}^{327}}
\sum_{\bk\in\RBZ}\sum_{\sigma}\sigma
\left[
\left\langle
\alpha^\dagger_{\bk\sigma}
\beta_{\bk+\bQ,\sigma}
\right\rangle
+
\left\langle
\alpha^\dagger_{\bk+\bQ,\sigma}
\beta_{\bk\sigma}
\right\rangle
\right],
\label{eq:lp327_m_bilinear}
\end{equation}
and the real order parameter is
\begin{equation}
\Delta_{\alpha\beta}
=
-\frac12
\left(g_{\alpha\beta}+g_{\alpha\beta,1}\right)
M_{\alpha\beta}^{327}.
\label{eq:lp327_gap_eq}
\end{equation}
The LP327 Hartree--Fock calculation is performed at fixed total filling
$n_{327}=6.0$ electrons per LP327 unit cell.  The chemical-potential shift
$\delta\mu$ is adjusted so that the occupation of the retained
$\alpha,\beta,\beta^\prime$ bands together with the noninteracting background
bands satisfies this filling constraint.
Thus $\beta^\prime$ influences the LP327 solution through the filling constraint,
while it carries no residual SDW vertex in the main-text calculation.

\subsection{Non-self-conjugate LP4310 case}

For LP4310 the active pair is $\alpha$ and $\beta^\prime$, and the
commensurate wave vector used in the calculation is
$\bQ=(2\pi/3,2\pi/3)$, so $3\bQ$ is reciprocal but
$-\bQ\equiv2\bQ$ is distinct from $\bQ$ in the folded representation. Define sector operators
\begin{equation}
\alpha_{m,\bk\sigma}\equiv\alpha_{\bk+m\bQ,\sigma},
\qquad
\beta^\prime_{m,\bk\sigma}\equiv\beta^\prime_{\bk+m\bQ,\sigma},
\qquad
m=0,1,2,
\label{eq:sector_operators}
\end{equation}
with sector indices understood modulo 3.  The normalization below uses
$N_{\rm full}=3N_{\RBZ}$, where $N_{\RBZ}$ is the number of folded-zone
momenta.  In the folded basis
\begin{equation}
\Psi^{4310}_{\bk\sigma}
=
\left(
\alpha_{0,\bk\sigma},
\beta^\prime_{0,\bk\sigma},
\alpha_{1,\bk\sigma},
\beta^\prime_{1,\bk\sigma},
\alpha_{2,\bk\sigma},
\beta^\prime_{2,\bk\sigma}
\right)^T,
\label{eq:lp4310_folded_basis}
\end{equation}
We write
\begin{equation}
\xi^\lambda_m(\bk)\equiv \xi^\lambda_{\bk+m\bQ},
\qquad
\lambda=\alpha,\beta^\prime .
\label{eq:lp4310_sector_dispersion}
\end{equation}
The static Hartree-Fock Hamiltonian used to build the transverse bubble is
\begin{equation}
\begingroup
\setlength{\arraycolsep}{3pt}
\mathcal H^{4310}_{\sigma}(\bk)
=
\begin{pmatrix}
\xi^\alpha_0-\delta\mu
& \sigma\Delta_0
& \rho_\alpha
& \sigma\Delta_+
& \rho_\alpha^*
& \sigma\Delta_-
\\
\sigma\Delta_0^*
& \xi^{\beta^\prime}_0-\delta\mu
& \sigma\Delta_-^*
& \rho_{\beta^\prime}
& \sigma\Delta_+^*
& \rho_{\beta^\prime}^*
\\
\rho_\alpha^*
& \sigma\Delta_-
& \xi^\alpha_1-\delta\mu
& \sigma\Delta_0
& \rho_\alpha
& \sigma\Delta_+
\\
\sigma\Delta_+^*
& \rho_{\beta^\prime}^*
& \sigma\Delta_0^*
& \xi^{\beta^\prime}_1-\delta\mu
& \sigma\Delta_-^*
& \rho_{\beta^\prime}
\\
\rho_\alpha
& \sigma\Delta_+
& \rho_\alpha^*
& \sigma\Delta_-
& \xi^\alpha_2-\delta\mu
& \sigma\Delta_0
\\
\sigma\Delta_-^*
& \rho_{\beta^\prime}
& \sigma\Delta_+^*
& \rho_{\beta^\prime}^*
& \sigma\Delta_0^*
& \xi^{\beta^\prime}_2-\delta\mu
\end{pmatrix}.
\label{eq:lp4310_hf_matrix}
\endgroup
\end{equation}
For the HF-RPA calculation of LP4310, the one-particle energies are multiplied
by a global bandwidth rescaling factor of $0.24$.  The fixed filling is taken
from the corresponding normal-state six-band model at the rescaled reference
chemical potential $\mu_0=-0.0161~{\rm eV}$.  In the ordered-state calculation
$\delta\mu$ is adjusted so that the retained $\alpha,\beta^\prime$ bands and
the noninteracting spectator bands together keep this total filling.
Equation~\eqref{eq:lp4310_hf_matrix} displays the active
$\alpha$--$\beta^\prime$ block.  When spectator Fermi-level bands are included
in the RPA bubble, they are added as diagonal folded one-particle sectors with
no static SDW fields and with zero residual interaction vertices.
The two leading SDW harmonics are independent:
\begin{align}
M_{\alpha\beta^\prime}^{(+)}
&=
\frac{1}{N_{\rm full}}
\sum_{\bk\in\RBZ}\sum_{m=0}^{2}\sum_{\sigma}
\sigma
\left\langle
\alpha^\dagger_{m+1,\bk\sigma}
\beta^\prime_{m,\bk\sigma}
\right\rangle,
\nonumber\\
M_{\alpha\beta^\prime}^{(-)}
&=
\frac{1}{N_{\rm full}}
\sum_{\bk\in\RBZ}\sum_{m=0}^{2}\sum_{\sigma}
\sigma
\left\langle
\alpha^\dagger_{m-1,\bk\sigma}
\beta^\prime_{m,\bk\sigma}
\right\rangle .
\label{eq:mpm_bilinears}
\end{align}
The fields obey
\begin{equation}
\begin{pmatrix}
\Delta_+\\
\Delta_-
\end{pmatrix}
=
-\frac12
\begin{pmatrix}
g_{\alpha\beta^\prime} & g_{\alpha\beta^\prime,1}\\
g_{\alpha\beta^\prime,1} & g_{\alpha\beta^\prime}
\end{pmatrix}
\begin{pmatrix}
M_{\alpha\beta^\prime}^{(+)}\\
M_{\alpha\beta^\prime}^{(-)}
\end{pmatrix}.
\label{eq:delta_pm_update}
\end{equation}

Beyond the leading $\pm\bQ$ order parameters, the nonlinear Hartree-Fock
equations can generate higher harmonics as the SDW amplitude increases~\cite{doi:10.1143/JPSJ.70.866}.
For the approximate commensurate wave vector $\bQ=(2\pi/3,2\pi/3)$, the third harmonic is folded
back to zero momentum, $3\bQ\equiv0$.  The corresponding interband SDW
component therefore connects $\alpha_m$ and $\beta^\prime_m$ within the same
folded sector:
\begin{equation}
M_{\alpha\beta^\prime}^{(0)}
=
\frac{1}{N_{\rm full}}
\sum_{\bk\in\RBZ}\sum_{m=0}^{2}\sum_{\sigma}
\sigma
\left\langle
\alpha^\dagger_{m,\bk\sigma}
\beta^\prime_{m,\bk\sigma}
\right\rangle ,
\label{eq:mzero}
\end{equation}
and is represented in Eq.~\eqref{eq:lp4310_hf_matrix} by the matrix element
$\Delta_0$,
\begin{equation}
\Delta_0
=
-\frac12\left[
g_{\alpha\beta^\prime}M_{\alpha\beta^\prime}^{(0)}
+g_{\alpha\beta^\prime,1}
\left(M_{\alpha\beta^\prime}^{(0)}\right)^*
\right].
\label{eq:delta_zero}
\end{equation}
Including $\Delta_0$ keeps the Hartree-Fock state and the RPA vertex tied to
the same set of residual interaction channels.

The induced intraband charge harmonics are
\begin{align}
\eta_\alpha
=
\frac{1}{N_{\rm full}}
\sum_{\bk\in\RBZ}\sum_{m=0}^{2}\sum_{\sigma}
\left\langle
\alpha^\dagger_{m+2,\bk\sigma}
\alpha_{m,\bk\sigma}
\right\rangle,
\nonumber\\
\eta_{\beta^\prime}
=
\frac{1}{N_{\rm full}}
\sum_{\bk\in\RBZ}\sum_{m=0}^{2}\sum_{\sigma}
\left\langle
\left(\beta^\prime_{m+2,\bk\sigma}\right)^\dagger
\beta^\prime_{m,\bk\sigma}
\right\rangle ,
\label{eq:eta_abp}
\end{align}
with single-particle fields
\begin{align}
\rho_\alpha
=
\frac{g_{\alpha\alpha}}{2}\eta_\alpha
+\left(g_{\alpha\beta^\prime}
-\frac{g_{\alpha\beta^\prime,2}}{2}\right)\eta_{\beta^\prime},
\nonumber\\
\rho_{\beta^\prime}
=
\frac{g_{\beta^\prime\beta^\prime}}{2}\eta_{\beta^\prime}
+\left(g_{\alpha\beta^\prime}
-\frac{g_{\alpha\beta^\prime,2}}{2}\right)\eta_\alpha .
\label{eq:rho_abp_update}
\end{align}

\section{Transverse RPA response}

The transverse response is evaluated in the ordered Hartree-Fock state.  For
an $N$-sector commensurate SDW, an external momentum $\bq$ couples the folded
transfer momenta
$\bq,\bq+\bQ,\ldots,\bq+(N-1)\bQ$.  For a projected band set $\mathcal B$, define
\begin{equation}
S^+_{m;l_1l_2}(\bq)
\equiv
\sum_{\bk\in\RBZ}\sum_{r=0}^{N-1}
c^\dagger_{l_1,\bk+\bq+(r+m)\bQ,\uparrow}
c_{l_2,\bk+r\bQ,\downarrow},
\label{eq:transverse_operator}
\end{equation}
where $l_i\in\mathcal B$, $m=0,\ldots,N-1$, and folded-sector indices are
understood modulo $N$.  We use
$N_{\rm full}=N N_{\RBZ}$, where $N_{\RBZ}$ is the number of momenta in the
folded Brillouin zone.
The corresponding four-index transverse susceptibility is defined by
\begin{equation}
\left[
\chi^{+-}_{m m'}(\bq,\ii\omega_n)
\right]^{l_1l_3}_{l_2l_4}
=
\int_0^\beta d\tau\,
\ee^{\ii\omega_n\tau}
\left\langle
T_\tau
S^+_{m;l_3l_4}(\bq,\tau)
S^-_{m';l_2l_1}(-\bq,0)
\right\rangle .
\label{eq:four_index_chi_definition}
\end{equation}

The bare ordered-state bubble is evaluated by diagonalizing the ordered
Hartree-Fock Hamiltonian in the spin-up and spin-down sectors and inserting the
corresponding eigenvectors into the particle-hole bubble.  We write
\begin{equation}
\mathcal H_\uparrow(\bk+\bq)u^{(\nu)}(\bk+\bq)
=
E_{\nu\uparrow}(\bk+\bq)u^{(\nu)}(\bk+\bq),
\label{eq:up_hf_eigenvector}
\end{equation}
for the spin-up sector, and
\begin{equation}
\mathcal H_\downarrow(\bk)v^{(\mu)}(\bk)
=
E_{\mu\downarrow}(\bk)v^{(\mu)}(\bk).
\label{eq:down_hf_eigenvector}
\end{equation}
for the spin-down sector.  Here $\nu$ and $\mu$ label Hartree-Fock
quasiparticle eigenstates.  The component
$u^{(\nu)}_{l,r}(\bk+\bq)$ is the amplitude of the $\nu$-th spin-up
eigenstate on band $l$ in folded sector $r$; $v^{(\mu)}_{l,r}(\bk)$ is the
corresponding spin-down amplitude.

Using these eigenvectors, the transverse bilinear
$c^\dagger_{l_1\uparrow}c_{l_2\downarrow}$ has the folded-sector matrix element
\begin{equation}
A^{\nu\mu}_{m;l_1l_2}(\bk,\bq)
=
\sum_{r=0}^{N-1}
\left[
u^{(\nu)}_{l_1,r+m}(\bk+\bq)
\right]^*
v^{(\mu)}_{l_2,r}(\bk).
\label{eq:transfer_amplitude}
\end{equation}
The Matsubara sum then gives
\begin{equation}
\left[
\chi^{+-}_{0}(\bq,\omega)
\right]^{l_1l_3}_{l_2l_4}(m,m')
=
-\frac{1}{N_{\rm full}}
\sum_{\bk\in\RBZ}\sum_{\nu,\mu}
\left[A^{\nu\mu}_{m;l_1l_2}(\bk,\bq)\right]^*
A^{\nu\mu}_{m';l_3l_4}(\bk,\bq)
\frac{
f(E_{\nu\uparrow}(\bk+\bq))-f(E_{\mu\downarrow}(\bk))
}{
\omega+E_{\nu\uparrow}(\bk+\bq)-E_{\mu\downarrow}(\bk)+\ii\eta
}.
\label{eq:folded_bubble_eigen}
\end{equation}

The residual vertex is local in folded transfer-sector space,
\begin{equation}
\left[
U^{+-}_{{\rm full};m m'}
\right]^{l_1l_3}_{l_2l_4}
=
\delta_{m m'}
\left[U^{+-}_{\rm pair}(\mathcal B)\right]^{l_1l_3}_{l_2l_4},
\label{eq:full_vertex}
\end{equation}
where $U^{+-}_{\rm pair}(\mathcal B)$ contains the active-sector two-band
vertex given in Eq.~\eqref{eq:upair_main}, while matrix
elements involving band pairs outside that sector are set to zero.  The
folded RPA ladder is
\begin{equation}
\hat\chi^{+-}_{\rm RPA}(\bq,\omega)
=
\left[
\hat 1
-\hat\chi^{+-}_0(\bq,\omega)\hat U^{+-}_{\rm full}
\right]^{-1}
\hat\chi^{+-}_0(\bq,\omega).
\label{eq:folded_rpa}
\end{equation}
Here the hats indicate matrices in the folded transfer indices and in the
four band indices of Eq.~\eqref{eq:four_index_chi_definition}.

It is useful to separate the active pair-channel subspace $\mathcal A$, where
the residual vertex is nonzero, from the remaining pair channels $\mathcal R$
that carry no residual vertex.  In this block notation,
\begin{equation}
\chi_0
=
\begin{pmatrix}
\chi_{0,\mathcal A\mathcal A} & \chi_{0,\mathcal A\mathcal R}\\
\chi_{0,\mathcal R\mathcal A} & \chi_{0,\mathcal R\mathcal R}
\end{pmatrix},
\qquad
\hat U_{\rm full}
=
\begin{pmatrix}
\hat U & 0\\
0 & 0
\end{pmatrix}.
\label{eq:rpa_active_spectator_blocks}
\end{equation}
Equation~\eqref{eq:folded_rpa} gives the spectator block
\begin{equation}
\chi^{\rm RPA}_{\mathcal R\mathcal R}
=
\chi_{0,\mathcal R\mathcal R}
+
\chi_{0,\mathcal R\mathcal A}\hat U
\left(
\hat 1_{\mathcal A}-\chi_{0,\mathcal A\mathcal A}\hat U
\right)^{-1}
\chi_{0,\mathcal A\mathcal R}.
\label{eq:spectator_rpa_block}
\end{equation}
Thus the additional Fermi-level bands in the LP4310 bubble add bare
particle-hole spectral weight and redistribute the RPA-enhanced weight through
the mixed bare-bubble blocks.  As a result, the transverse spectrum summed over
all Fermi-level bands is broader than the spectrum projected only onto the
interacting $\alpha$--$\beta^\prime$ pair.

The physical response at the external momentum $\bq$ is obtained, in the
folded representation, by selecting the block with both folded transfer
indices equal to zero:
\begin{equation}
\left[\chi^{+-}_{\rm phys}(\bq,\omega)\right]^{l_1l_3}_{l_2l_4}
=
\left[
\chi^{+-}_{{\rm RPA};00}(\bq,\omega)
\right]^{l_1l_3}_{l_2l_4},
\label{eq:physical_block}
\end{equation}
The other folded blocks describe the coupling to $\bq+m\bQ$ generated by the
SDW order.

The plotted intra- and interband responses are obtained from
Eq.~\eqref{eq:physical_block} by specifying the band labels included in the
final contraction. For a subset $\Lambda$ of the band set used in the RPA bubble, we use
\begin{align}
S^+_{\rm intra}(\bq;\Lambda)
&=
\sum_{\lambda\in\Lambda}
S^+_{0;\lambda\lambda}(\bq),
&
S^+_{\rm inter}(\bq;\Lambda)
&=
\sum_{\substack{\lambda_1,\lambda_2\in\Lambda\\
\lambda_1\ne\lambda_2}}
S^+_{0;\lambda_1\lambda_2}(\bq),
\label{eq:projected_transverse_bilinears}
\end{align}
with the lowering operators defined by Hermitian conjugation.  Expanding the
corresponding projected correlators in Eq.~\eqref{eq:physical_block} gives
\begin{align}
\chi^{+-}_{\rm intra}(\bq,\omega;\Lambda)
&=
\sum_{\lambda,\lambda'\in\Lambda}
\left[
\chi^{+-}_{\rm phys}(\bq,\omega)
\right]^{\lambda\lambda'}_{\lambda\lambda'} ,
\label{eq:intra_projection}
\end{align}
and
\begin{align}
\chi^{+-}_{\rm inter}(\bq,\omega;\Lambda)
&=
\sum_{\substack{
\lambda_1,\lambda_2,\lambda_3,\lambda_4\in\Lambda\\
\lambda_1\ne\lambda_2,\ \lambda_3\ne\lambda_4}}
\left[
\chi^{+-}_{\rm phys}(\bq,\omega)
\right]^{\lambda_1\lambda_3}_{\lambda_2\lambda_4}.
\label{eq:inter_projection}
\end{align}
The total plotted response is
\begin{equation}
\chi^{+-}_{\rm total}(\bq,\omega;\Lambda)
=
\chi^{+-}_{\rm intra}(\bq,\omega;\Lambda)
+
\chi^{+-}_{\rm inter}(\bq,\omega;\Lambda).
\label{eq:intra_inter_total_projection}
\end{equation}
For LP327, the folded problem has $N=2$.  The RPA bubble and the final
contraction use the three low-energy bands
$\Lambda=\{\alpha,\beta,\beta^\prime\}$, while the residual interaction vertex
is nonzero only in the $\alpha$--$\beta$ block.  For LP4310, the folded problem
has $N=3$.  The residual interaction vertex is nonzero only in the active
$\alpha$--$\beta^\prime$ sector, while the RPA bubble also includes the
additional bands crossing the Fermi level.  The final contraction therefore
uses $\Lambda$ equal to the full set of Fermi-level bands included in the
LP4310 RPA calculation.  In both cases the contraction is performed after the
full folded RPA problem has been solved.

\section{Orbital and layer projection of spin and charge textures}

Local spin and charge textures are obtained by projecting the ordered-state
density matrix to the microscopic orbital-layer basis.  For LP327, let
$\bk_0=\bk$ and $\bk_1=\bk+\bQ$, and write the band-to-orbital eigenvectors
of the three retained Fermi-level bands as $u_\alpha(\bk_s)$,
$u_\beta(\bk_s)$, and $u_{\beta^\prime}(\bk_s)$, with $s=0,1$.  If an
eigenvector of Eq.~\eqref{eq:lp327_hf_matrix} has amplitudes
$A_{\nu s},B_{\nu s},B^\prime_{\nu s}$ in the folded retained-band basis,
it is represented in the doubled LP327 orbital basis
$(C^{327}_{\bk_0\sigma},C^{327}_{\bk_1\sigma})$ as
\begin{equation}
\widetilde\varphi^{327}_{\nu\sigma}(\bk)
=
\begin{pmatrix}
A_{\nu0}u_\alpha(\bk_0)
+B_{\nu0}u_\beta(\bk_0)
+B^\prime_{\nu0}u_{\beta^\prime}(\bk_0)\\
A_{\nu1}u_\alpha(\bk_1)
+B_{\nu1}u_\beta(\bk_1)
+B^\prime_{\nu1}u_{\beta^\prime}(\bk_1)
\end{pmatrix}.
\label{eq:embedded_lp_vector}
\end{equation}
The retained-subspace density matrix has $8\times8$ blocks,
\begin{equation}
\widetilde\rho^{327}_\sigma(\bk)
=
\sum_\nu f(E_{\nu\sigma}(\bk))
\widetilde\varphi^{327}_{\nu\sigma}(\bk)
\left[\widetilde\varphi^{327}_{\nu\sigma}(\bk)\right]^\dagger
=
\begin{pmatrix}
\rho_\sigma^{00} & \rho_\sigma^{01}\\
\rho_\sigma^{10} & \rho_\sigma^{11}
\end{pmatrix}.
\label{eq:folded_density_matrix}
\end{equation}
To make the orbital projection definite, we fix the remaining phase freedom of
the $\alpha$ and $\beta$ Bloch states by their dominant mirror-resolved
$d_{x^2-y^2}$ projections.  In the LP327 orbital basis $C^{327}_{\bk\sigma}$,
define
\begin{equation}
\left|p^{327}_{x,+}\right\rangle
=
\frac{1}{2}(1,0,1,0,1,0,1,0)^T,\qquad
\left|p^{327}_{x,-}\right\rangle
=
\frac{1}{2}(1,0,1,0,-1,0,-1,0)^T ,
\label{eq:lp327_phase_vectors}
\end{equation}
where $+$ and $-$ denote even and odd mirror parity.  We choose the phases such
that
\begin{equation}
\left\langle p^{327}_{x,+}\middle|u_{\alpha}(\bk_s)\right\rangle
\in\mathbb R_{>0},
\qquad
\left\langle p^{327}_{x,-}\middle|u_{\beta}(\bk_s)\right\rangle
\in\mathbb R_{>0},
\qquad
s=0,1 .
\label{eq:lp327_phase_fixing}
\end{equation}
The intracell positions entering the LP327 density reconstruction are
\begin{equation}
\boldsymbol\tau_A=(0,0),\qquad
\boldsymbol\tau_B=(1/2,1/2),
\label{eq:lp327_tau}
\end{equation}
for both layers and for both $d_{x^2-y^2}$ and $d_{z^2}$ orbitals.  For
orbital-layer index $a=(i,\eta)$, with $\eta=x,z$ and
$\boldsymbol\tau_a=\boldsymbol\tau_i$, the density in cell $\bR$ is evaluated with
$N_{\rm full}^{327}=2N_{\RBZ}$ and $\bk\in\RBZ$:
\begin{equation}
n_{a\sigma}(\bR)
=
\frac{1}{N_{\rm full}^{327}}
\sum_{\bk\in\RBZ}
\left[
\rho^{00}_{\sigma;aa}(\bk)
+\rho^{11}_{\sigma;aa}(\bk)
+\ee^{\ii\bQ\cdot(\bR+\boldsymbol\tau_a)}
\rho^{01}_{\sigma;aa}(\bk)
+\ee^{-\ii\bQ\cdot(\bR+\boldsymbol\tau_a)}
\rho^{10}_{\sigma;aa}(\bk)
\right].
\label{eq:lp327_density_reconstruction}
\end{equation}
For the ordering vector $\bQ=(0,\pi)$, the corresponding reference-cell phase
factors are $\ee^{\ii\bQ\cdot\boldsymbol\tau_A}=1$ and
$\ee^{\ii\bQ\cdot\boldsymbol\tau_B}=\ii$.
The projected local moment on Ni site $i$ is
\begin{equation}
m_i(\bR)
=
\sum_{\eta=x,z}
\left[
n_{(i,\eta),\uparrow}(\bR)-n_{(i,\eta),\downarrow}(\bR)
\right].
\label{eq:local_moment}
\end{equation}
For LP4310, let $\bk_m=\bk+m\bQ$, with $m=0,1,2$ understood modulo 3, and let
$u_{\alpha}(\bk_m)$ and $u_{\beta^\prime}(\bk_m)$ be the corresponding
six-component orbital-layer eigenvectors in the basis of
Eq.~\eqref{eq:lp4310_basis}.  If an eigenvector of
Eq.~\eqref{eq:lp4310_hf_matrix} has amplitudes $A_{\nu m}$ and $B_{\nu m}$
on $\alpha_m$ and $\beta^\prime_m$, respectively, it is represented in the
folded orbital-layer basis as
\begin{equation}
\widetilde\varphi^{4310}_{\nu\sigma}(\bk)
=
\begin{pmatrix}
A_{\nu0}u_{\alpha}(\bk_0)+B_{\nu0}u_{\beta^\prime}(\bk_0)\\
A_{\nu1}u_{\alpha}(\bk_1)+B_{\nu1}u_{\beta^\prime}(\bk_1)\\
A_{\nu2}u_{\alpha}(\bk_2)+B_{\nu2}u_{\beta^\prime}(\bk_2)
\end{pmatrix}.
\label{eq:lp4310_embedded_vector}
\end{equation}
The corresponding folded density matrix is
\begin{equation}
\widetilde\rho^{4310}_{\sigma}(\bk)
=
\sum_\nu
f(E_{\nu\sigma}(\bk))
\widetilde\varphi^{4310}_{\nu\sigma}(\bk)
\left[\widetilde\varphi^{4310}_{\nu\sigma}(\bk)\right]^\dagger
=
\left(\rho^{rs}_{\sigma}(\bk)\right)_{r,s=0,1,2},
\label{eq:lp4310_folded_density}
\end{equation}
where each $\rho^{rs}_{\sigma}$ is a $6\times6$ matrix in the orbital-layer
basis of Eq.~\eqref{eq:lp4310_basis}.

For the layer-resolved projection we use the same mirror-resolved
$d_{x^2-y^2}$ phase convention as in the LP327 case.  Define
\begin{equation}
\left|p^{4310}_{x,+}\right\rangle
=
\frac{1}{\sqrt3}(1,0,1,0,1,0)^T,
\qquad
\left|p^{4310}_{x,-}\right\rangle
=
\frac{1}{\sqrt2}(1,0,0,0,-1,0)^T .
\label{eq:lp4310_phase_vectors}
\end{equation}
The phases of the retained $\alpha$ and $\beta^\prime$ Bloch states are chosen
such that
\begin{equation}
\left\langle p^{4310}_{x,+}\middle|u_{\alpha}(\bk_m)\right\rangle
\in\mathbb R_{>0},
\qquad
\left\langle p^{4310}_{x,-}\middle|u_{\beta^\prime}(\bk_m)\right\rangle
\in\mathbb R_{>0},
\qquad
m=0,1,2 .
\label{eq:lp4310_phase_fixing}
\end{equation}

For the three-sector LP4310 projection,
$N_{\rm full}^{4310}=3N_{\RBZ}$, and all folded momentum sums below are over
$\bk\in\RBZ$.  For orbital-layer index $a=(\ell,\eta)$, define the
spin-resolved density harmonics
\begin{equation}
n^{4310}_{a\sigma}(p\bQ)
=
\frac{1}{N_{\rm full}^{4310}}
\sum_{\bk\in\RBZ}\sum_{m=0}^{2}
\rho^{m+p,m}_{\sigma;aa}(\bk),
\qquad
p=1,2,
\label{eq:lp4310_spin_charge_projection}
\end{equation}
with sector indices understood modulo 3.  The layer-resolved spin and charge
quantities are obtained by summing over the two orbitals on each layer,
\begin{equation}
m_{\ell}(\bQ)
=
\mathrm{Re}\sum_{a\in\ell}
\left[
n^{4310}_{a\uparrow}(\bQ)
-
n^{4310}_{a\downarrow}(\bQ)
\right],
\qquad
\delta n_{\ell}(2\bQ)
=
2\left|
\sum_{a\in\ell}
\left[
n^{4310}_{a\uparrow}(2\bQ)
+
n^{4310}_{a\downarrow}(2\bQ)
\right]
\right|.
\label{eq:lp4310_layer_projection}
\end{equation}
Since $n^{4310}_{a\uparrow}(2\bQ)+n^{4310}_{a\downarrow}(2\bQ)$ is built from
intraband charge coherences, the induced $2\bQ$ charge modulation is mirror
even, while the primary interband SDW between opposite mirror sectors is
mirror odd.

\section{Variational check of the large-$J_\perp$ bilayer Heisenberg limit}

The main text argues that the large interlayer exchange scale extracted from
local-spin fits should not be interpreted directly as a microscopic
superexchange coupling in a weak-moment metal.  We therefore use a quantum
spin-$1/2$ bilayer Heisenberg model as a local-moment benchmark for this
large-$J_\perp$ interpretation.

The relevant strong-coupling limit is already apparent for an isolated
vertical interlayer dimer, or rung,
\begin{equation}
H_{\perp,i}
=
J_\perp \mathbf S_{1,i}\cdot\mathbf S_{2,i},
\qquad
J_\perp>0,
\end{equation}
the spin-$1/2$ ground state is the singlet
$|s_i\rangle=(|\uparrow\downarrow\rangle-|\downarrow\uparrow\rangle)/\sqrt2$,
with energy $-3J_\perp/4$, while the triplet states lie higher by $J_\perp$.
Thus a large antiferromagnetic $J_\perp$ first favors an interlayer singlet.
In-plane exchanges can compete only by admixing triplet configurations.
Static magnetic order, including a double-stripe pattern, therefore requires
triplet weight to condense out of the rung-singlet background and is not
expected when $J_\perp$ is much larger than the in-plane exchange scales.

We now apply this local-moment benchmark to the exchange hierarchy inferred in
Ref.~\cite{chen2026naturemagnetismbilayernickelate}.  Following the
spin--spinless stripe model of that work, we normalize the interlayer coupling
to $J_\perp=1$, so all energies quoted below are in units of $J_\perp$, and use
\begin{equation}
J_{1x}=0.0931,\qquad
J_{1y}=0.0592,\qquad
J_2=0.1162,\qquad
J_\perp=1 .
\label{eq:vmc_exchange_ratios}
\end{equation}
The spin Hamiltonian is
\begin{align}
H_{\rm spin}
&=
J_{1x}\sum_{\langle ij\rangle_x,l}
\mathbf S_{l,i}\cdot\mathbf S_{l,j}
+J_{1y}\sum_{\langle ij\rangle_y,l}
\mathbf S_{l,i}\cdot\mathbf S_{l,j}
\nonumber\\
&\quad
+J_2\sum_{\langle\langle ij\rangle\rangle,l}
\mathbf S_{l,i}\cdot\mathbf S_{l,j}
+J_\perp\sum_i
\mathbf S_{1,i}\cdot\mathbf S_{2,i}.
\label{eq:bilayer_spin_model}
\end{align}
Here $l=1,2$ labels the two layers and $i=(x_i,y_i)$ labels an in-plane
site.  The operators $\mathbf S_{l,i}$ are spin-$1/2$ operators.  The sums
$\langle ij\rangle_x$ and $\langle ij\rangle_y$ run over nearest-neighbor
bonds along the two in-plane directions within each layer, while
$\langle\langle ij\rangle\rangle$ runs over the two diagonal
next-nearest-neighbor bonds.  The last term couples the two sites on the same
vertical bond.  Positive exchange constants denote antiferromagnetic couplings.

The variational state is a Gutzwiller-projected fermionic parton state,
\begin{equation}
|\Psi_{\rm VMC}\rangle=P_G|\Phi_{\rm MF}\rangle ,
\label{eq:vmc_projected_state}
\end{equation}
where $|\Phi_{\rm MF}\rangle$ is the ground state of a bilayer mean-field
Hamiltonian.  This mean-field Hamiltonian allows intralayer hopping channels
$\chi_{1x},\chi_{1y},\chi_2$, intralayer singlet-pairing channels
$\eta_{1x},\eta_{1y},\eta_2$, interlayer hopping and pairing channels
$\chi_\perp,\eta_\perp$, and a stripe magnetic field $m_{\rm stripe}$.
Schematically,
\begin{equation}
H_{\rm MF}=H_\chi+H_\eta+H_m ,
\label{eq:vmc_mf_hamiltonian}
\end{equation}
where $H_\chi$ and $H_\eta$ collect the hopping and singlet-pairing terms on
the same bonds as Eq.~\eqref{eq:bilayer_spin_model}, including the vertical
interlayer bond.  The magnetic term is
\begin{equation}
H_m
=
\frac12
\sum_{l,i,\sigma}
\sigma(-1)^{l-1}
m_{\rm stripe}(-1)^{x_i}
f^\dagger_{l i\sigma}f_{l i\sigma}.
\label{eq:vmc_magnetic_fields}
\end{equation}
The variational parameters are optimized before the final measurements.

After optimization, the variational states remain close to this rung-singlet
limit.  The stripe field relaxes to zero within numerical accuracy,
$m_{\rm stripe}\to0$, and the rung correlation
$\langle\mathbf S_{1,i}\cdot\mathbf S_{2,i}\rangle$ remains close to the
isolated-singlet value $-3/4$ for all system sizes considered, as summarized
in Table~\ref{tab:vmc_rung_results}.

\begin{table}[htbp]
\centering
\begin{tabular}{ccc}
\hline
$L_x\times L_y$ & $E/N_{\rm site}$ &
$\langle\mathbf S_{1,i}\cdot\mathbf S_{2,i}\rangle$\\
\hline
$10\times10$ & $-0.381926\pm7\times10^{-6}$ & $-0.73713\pm8.5\times10^{-5}$\\
$12\times12$ & $-0.381921\pm6.5\times10^{-6}$ & $-0.73710\pm8.7\times10^{-5}$\\
$14\times14$ & $-0.381924\pm5\times10^{-6}$ & $-0.73713\pm6.9\times10^{-5}$\\
$16\times16$ & $-0.381923\pm3.7\times10^{-6}$ & $-0.73713\pm6.9\times10^{-5}$\\
\hline
\end{tabular}
\caption{VMC estimates for the bilayer Heisenberg model in
Eq.~\eqref{eq:bilayer_spin_model}, using the exchange ratios in
Eq.~\eqref{eq:vmc_exchange_ratios}.  Energies are measured in units of
$J_\perp$.}
\label{tab:vmc_rung_results}
\end{table}

\begin{figure}[htbp]
\centering
\includegraphics[width=0.75\textwidth]{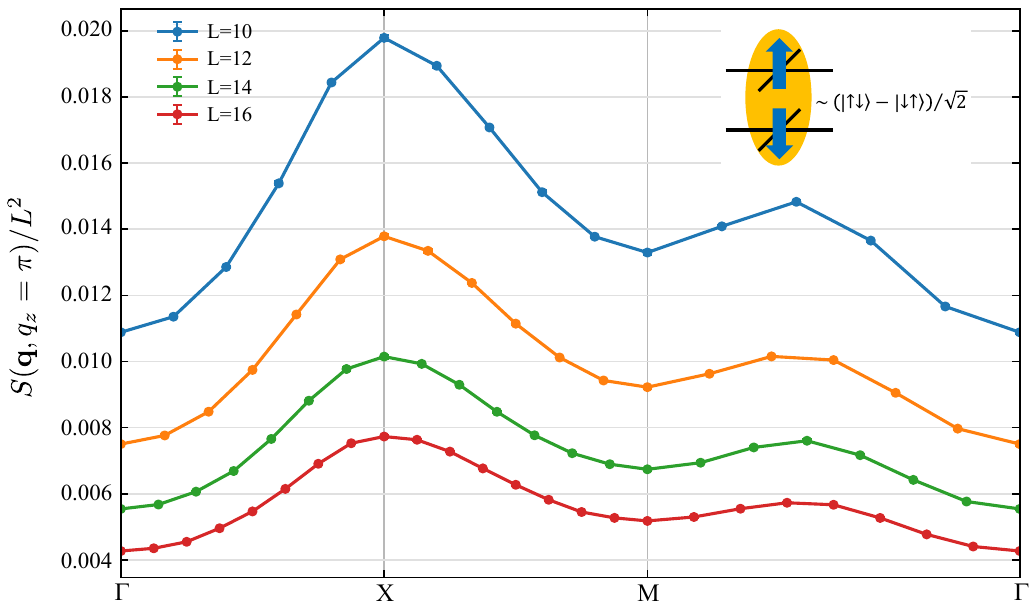}
\caption{Static spin structure factor
$S(\mathbf q,q_z=\pi)/L^2$ for the optimized variational states of
Hamiltonian~\eqref{eq:bilayer_spin_model}, plotted along
$\Gamma-X-M-\Gamma$.  The decrease with increasing system size is consistent
with the absence of robust magnetic long-range order in this large-$J_\perp$
local-moment benchmark. The upper-right schematic illustrates the
interlayer spin singlet on a vertical dimer.}
\label{fig:vmc_structure_factor}
\end{figure}

The static spin structure factor gives an independent check.  As shown in
Fig.~\ref{fig:vmc_structure_factor}, $S(\mathbf q,q_z=\pi)/L^2$ rapidly
decreases with increasing system size along the $\Gamma-X-M-\Gamma$ path,
rather than developing a size-stable magnetic peak.

Taken together, the optimized variational parameters, rung correlations, and
finite-size behavior of the structure factor indicate that this local-moment
model lies on the interlayer-dimer side rather than in a weak-moment
stripe-ordered phase.  This supports the main-text interpretation that the
large exchange scale appearing in phenomenological spin-wave fits should not
be identified directly with a microscopic superexchange coupling of the
itinerant SDW metal.



\end{document}